\documentclass[prd, twocolumn, tightenlines, twoside, secnumarabic, superscriptaddress, preprintnumbers, nofootinbib, notitlepage]{revtex4-1}
\usepackage{graphicx}
\usepackage{epstopdf}
\usepackage{amsmath}
\usepackage{amsfonts}
\usepackage{amssymb}
\usepackage{appendix}
\usepackage{comment}
\usepackage{color}
\usepackage{slashed}
\usepackage{caption}
\usepackage{subcaption}
\usepackage{setspace}
\usepackage{footnote}
\usepackage{multirow}
\usepackage{mathrsfs}
\usepackage{hyperref}
\usepackage{cleveref}
\usepackage{wrapfig}

\definecolor{niceblue}{rgb}{0.388235, 0.627451, 0.847059}
\definecolor{nicered}{rgb}{0.7,0.1,0.1}
\definecolor{nicegreen}{rgb}{0.1,0.5,0.1}
\hypersetup{colorlinks,citecolor= nicegreen,linkcolor= nicered}

\newcommand{\be}{\begin{equation}}
\newcommand{\ee}{\end{equation}}
\newcommand{\ba}{\begin{array}}
\newcommand{\ea}{\end{array}}
\newcommand{\bea}{\begin{eqnarray}}
\newcommand{\eea}{\end{eqnarray}}
\newcommand{\balg}{\begin{align}}
\newcommand{\ealg}{\end{align}}
\newcommand{\bit}{\begin{itemize}}
\newcommand{\eit}{\end{itemize}}
\newcommand{\trm}[1]{\textrm{#1}}

\definecolor{JMBcomment}{RGB}{64, 214, 136}

\newcommand{\Mpc}{\trm{\Mpc}}
\newcommand{\yr}{\trm{\yr}}
\newcommand{\eV}{\trm{\eV}}

\begin{document}

\singlespacing
\allowdisplaybreaks

\title{Future searches for light sterile neutrinos at nuclear reactors}

\author{Jeffrey M. Berryman}
\email{jeffberryman@berkeley.edu}

\affiliation{Department of Physics and Astronomy, University of Kentucky, Lexington, Kentucky 40506, USA}
\affiliation{Department of Physics, University of California, Berkeley, California 94720, USA}

\author{Luis A. Delgadillo}
\email{luisd@vt.edu}

\author{Patrick Huber}
\email{pahuber@vt.edu}

\affiliation{Center for Neutrino Physics, Department of Physics, Virginia Tech, Blacksburg, Virginia 24061, USA}

\preprint{N3AS-21-004}


\begin{abstract}
We study the optimization of a green-field, two-baseline reactor experiment with respect to the sensitivity for electron antineutrino disappearance in search of a light sterile neutrino. We consider both commercial and research reactors and identify as key factors the distance of closest approach and detector energy resolution. We find that a total of 5\,tons of detectors deployed at a commercial reactor with a closest approach of 25\,m can probe the mixing angle $\sin^22\theta$ down to $\sim5\times10^{-3}$ around $\Delta m^2\sim 1\,\mathrm{eV}^2$. The same detector mass deployed at a research reactor can be sensitive up to  $\Delta m^2\sim20-30\,\mathrm{eV}^2$ assuming a closest approach of 3\,m and excellent energy resolution, such as that projected for the Taishan Antineutrino Observatory. We also find that lithium doping of the reactor could be effective in increasing the sensitivity for higher $\Delta m^2$ values.
\end{abstract}


\maketitle

\section{Introduction}
\label{sec:intro}
\setcounter{equation}{0}

Neutrino physics as an experimental science started with the discovery of Cowan and Reines using neutrinos from a reactor in 1956~\cite{Cowan:1992xc}. Since then, reactor measurements have been a mainstay in the quest to understand neutrino properties. KamLAND~\cite{Eguchi:2002dm}, Double Chooz~\cite{Abe:2011fz}, Daya Bay~\cite{An:2012eh} and RENO~\cite{Ahn:2012nd} have played central roles in establishing the oscillation of the three active neutrino flavors. In the near future, the JUNO experiment will provide the best measurement of the so-called solar parameters $\theta_{12}$ and $\Delta m^2_{21}$ as well as determine the mass ordering~\cite{An:2015jdp}. In the run-up to the 2011/2012 measurement of $\theta_{13}$, the question of how to predict the reactor antineutrino spectrum received renewed attention~\cite{Mueller:2011nm,Huber:2011wv} and the surprising result was an upward correction to the predicted inverse beta decay (IBD) event rate by approximately 6\%. This, in turn, gave rise to the so-called reactor antineutrino anomaly (RAA)~\cite{Mention:2011rk}, which would be naturally explained if the $\bar\nu_e$ mixed with an additional species of neutrino with a mixing angle $\sin^22\theta\simeq0.1$ and $\Delta m^2\geq 1\,\mathrm{eV}^2$. For a review of the status of the field at that time, see Ref.~\cite{Abazajian:2012ys}. Enormous experimental progress has been made since then and  we refer the reader to the global fitting literature for details, see, e.g., Ref.~\cite{Giunti:2017yid, Dentler:2017tkw, Dentler:2018sju, Giunti:2019qlt, Berryman:2020agd,Giunti:2020uhv, Giunti:2021kab}; for a more experiment-centered review, see Ref.~\cite{Boser:2019rta}.

As far as reactor neutrino experiments are concerned, modern experiments (i.e., those conducted after 2011) \cite{NEOS:2016wee, DANSS:2018fnn, Serebrov:2020kmd, PROSPECT:2020sxr, STEREO:2020hup} all rely on a comparison of measured event rate spectra at different baselines and are thus \emph{independent} of flux predictions. Nonetheless, reactor flux predictions have been subject to intense study \cite{Hayes:2013wra,Hayen:2018uyg,Estienne:2019ujo,Hayen:2019eop}; see also Refs.~\cite{DayaBay:2017jkb, Giunti:2017nww, Giunti:2017yid, Gebre:2017vmm, Dentler:2017tkw, RENO:2018pwo, Giunti:2019qlt, Berryman:2019hme, Berryman:2020agd, Giunti:2021kab} . Moreover, the rate anomaly has not, as yet, been resolved, though it may be less significant than originally suggested.\footnote{Recent work~\cite{Kopeikin:2021ugh}, if confirmed, would provide a simple and consistent solution to the RAA by shifting the ratio of $^{239}$Pu to $^{235}$U in the integrated beta spectrum. In this case, beta and neutrino data would be in good agreement with both summation and conversion calculations of the reactor flux.} The spectral anomaly known as the 5-MeV bump \cite{DoubleChooz:2014kuw, DayaBay:2015lja, RENO:2015ksa, PROSPECT:2020sxr, STEREO:2020hup} is entirely unresolved; solutions grounded in nuclear physics have been proposed (see, e.g., Refs.~\cite{Hayes:2015yka, Sonzogni:2017wxy, Mention:2017dyq, Littlejohn:2018hqm, Hayen:2019eop}) and it is likely not due to new physics~\cite{Berryman:2018jxt}. The experiments we will discuss here will contribute data on neutrino fluxes, but this is not a focus of our work. Instead, we ask what the best possible reactor neutrino experiment would look like to either find a light sterile neutrino or to exclude a sizable mixing. We also focus exclusively on electron (anti)neutrinos and their potential mixing with a sterile neutrino. This choice is motivated, to some degree, by the fact that all data on $\nu_e\to \nu_e$ and $\bar\nu_e\to \bar\nu_e$ at this point are mutually consistent. The same cannot be said of the $\nu_\mu \to \nu_e$ and $\overline\nu_\mu \to \overline\nu_e$ data sets; additionally, the global disappearance data are known to be inconsistent with the global appearance data when interpreted in the context of a truly sterile fourth neutrino \cite{Dentler:2018sju, Boser:2019rta, Moulai:2019gpi, Adamson:2020jvo}. On top of all of this, an eV-scale sterile neutrino of the sort indicated by terrestrial oscillation anomalies is severely constrained by cosmological observations \cite{Berryman:2019nvr, Gariazzo:2019gyi, Hagstotz:2020ukm, Adams:2020nue}. Our hope is that a careful study of neutrino disappearance at reactors will facilitate future analyses of the global neutrino data set.

The reason nuclear reactors are the electron neutrino\footnote{This paper deals only with electron flavor neutrinos and/or antineutrinos and we rely on context to disambiguate these two cases.} source of choice is threefold: 
\begin{enumerate}
    \item They are free, in the sense that they are constructed for purposes other than neutrino physics.
    \item They are very bright, with experimentally accessible fluxes of up to $10^{13}\,\mathrm{cm}^{-2}\,\mathrm{s}^{-1}$.
    \item They produce antineutrinos which can be cleanly detected via IBD, which both provides a flavor tag and has a large (relative to other weak-interaction processes) cross section of approximately $6\times10^{-43}\,\mathrm{cm}^2$ per fission.
    \end{enumerate}
Other sources of electron neutrinos which have been considered for sterile neutrino searches include radioactive sources~\cite{GALLEX:1994rym, Abdurashitov:1996dp, GALLEX:1997lja, SAGE:1998fvr, Abdurashitov:2005tb, Grieb:2006mp, Kaether:2010ag, Borexino:2013xxa},\footnote{Since this work first appeared, the BEST Collaboration has reported a $\sim20$\% deficit in the antineutrino rate from a chromium source relative to prediction \cite{Barinov:2021asz}. These results can be interpreted \cite{Barinov:2021mjj} as a $\sim5\sigma$ indication of the existence of a sterile neutrino. It would be premature to declare that a sterile neutrino has been discovered, however, given the tension between this result and solar neutrino data \cite{Goldhagen:2021kxe}. This underscores the importance of fully addressing the anomalies in the reactor sector in order to form a coherent picture of $\nu_e$/$\overline{\nu}_e$ disappearance.} beta-beams~\cite{NeutrinoFactory:2004odt, Agarwalla:2009em}, kaon beams~\cite{Delgadillo:2020uvm}, beam-driven beta-decay sources~\cite{Conrad:2013sqa} and stored muon beams~\cite{Adey:2014rfv}. None of these other electron neutrino sources shares all of the advantages of nuclear reactors --- in particular they all would be quite costly and require detectors much larger than what will be considered here. On the other hand, some of these approaches like beam-driven beta-decay sources and stored muon beams would yield far better sensitivities than possible at a reactor. But the question remains: what sensitivities are a \emph{possible} with a reactor source? This is the subject of the present study.

\section{Green-Field studies}
\label{sec:gf}
\setcounter{equation}{0}

For the neutrino energies available at nuclear reactors and for the baselines that we will consider (less than 100\,m), oscillations among the three neutrinos of the Standard Model (SM) are negligible. Therefore, if a sterile neutrino\footnote{We will use the term ``sterile'' neutrino to refer to a generic fourth neutrino that participates in oscillations but not in weak interactions. This state may have interactions beyond those of the SM, but if these do not affect oscillations, then there is no need to distinguish it from a truly sterile neutrino.} exists and participates in oscillations, then any oscillations observed over these distances would be attributable to the new state. We may describe these oscillations in the two-flavor limit, writing the $\overline\nu_e$ survival probability as
\begin{equation}
    P_{\overline e \overline e} = 1 - \sin^2 2\theta \sin^2 \left( \frac{\Delta m^2 L}{4 E_\nu} \right).
    \label{eq:prob}
\end{equation}
Here, $L$ is the baseline over which the neutrinos propagate and $E_\nu$ is their energy. The parameter $\Delta m^2$ is the difference in the squares of the masses of two neutrino mass eigenstates; $\sin^2 2\theta$ is the mixing angle describing the amplitude of the oscillations. Probing sterile neutrinos amounts to determining $\sin^2 2\theta$ and $\Delta m^2$.

The overarching purpose of this study is to determine the optimal configuration(s) for a two-baseline reactor neutrino experiment with realistic operational assumptions. We categorize these by the class of reactor at which they are conducted --- at either a \emph{commercial} or a \emph{research} reactor facility. The salient difference between these is that the former is much larger ($R\sim2$ m) than the latter ($R\sim20$ cm); commercial cores are also more powerful and produce a larger flux of antineutrinos. However, the layout of the facilities at which these are housed prevents an experiment from operating within a certain distance of the core, typically of the order 10-25 m. Research cores, though less powerful, are less constrained in this regard: one can operate an experiment as close as 3-6 m from the core. The increase in flux from the shorter allowed baseline partially offsets the lower absolute rate of antineutrino production. We will consider each of these scenarios in turn.

We note similar previous works presented in Refs.~\cite{Heeger:2012tc, Heeger:2013ema}. Our work differs from these in two primary ways. Firstly, our focus is to systematically study the optimal physical configuration of the two detectors, depending on the type of reactor core. In contrast, these references are oriented more towards understanding the effects of, e.g., backgrounds and systematic uncertainties in dictating sensitivities. Secondly, our work benefits from nearly decade of additional experimental effort. In particular, these references largely predate measurements of spectral ratios \cite{NEOS:2016wee, DANSS:2018fnn, RENO:2018dro, DayaBay:2018yms, DoubleChooz:2019qbj, Serebrov:2020kmd, PROSPECT:2020sxr, STEREO:2020hup}, which favor decidedly different regions of the sterile neutrino parameter space than the traditional RAA. Our work serves as a complement to these early efforts.

\subsection{Methodology}
\label{sec:methods}

We consider pseudoexperiments in which the spectrum of reactor antineutrinos is observed at two baselines. We treat the detector as pointlike, but the reactor core has finite physical extent; we treat the latter as a perfect sphere. Because of this, we cannot use the oscillation probability in Eq.~\eqref{eq:prob} as is --- we must flux-average $P_{\overline e \overline e}$ over the production region. The effect of this is to average out high-frequency oscillations: if the oscillation wavelength is comparable to or smaller than the size of the production region, then the experiment will have muted sensitivity to the associated value of $\Delta m^2$. Details on this procedure and the resulting average survival probability are given in the Appendix.

The analysis window is taken to be $E_\text{prompt} \subset [2.0, \, 8.0]$ MeV; we generally assume a bin size of 250 keV, but occasionally consider finer spacing. We assume a constant, nonzero background in each energy bin, ignoring any systematic uncertainty on this background and assuming that the near and far baselines experience the same background rate per unit exposure.\footnote{In truth, the accidental background rate ought to be higher closer to the reactor core. We are ignoring this for simplicity, but note that this would require a more careful study under real-world conditions. Given the commensurate increase in signal rate, however, we do not expect this to dramatically change our findings.} We consider two energy response models as limiting cases of the detector resolution. The first is the response model of the PROSPECT experiment, which the collaboration has provided in the Supplemental Material to Ref.~\cite{Andriamirado:2020erz}. The response may be roughly described as a Gaussian with width $\sim7\%/\sqrt{E/\text{ [MeV]}}$, though it includes a non-Gaussian tail down to low energies. On the other end of the spectrum is the upcoming Taishan Antineutrino Observatory (TAO) experiment \cite{Abusleme:2020bzt}, for which the resolution will be of the order $\sim1\%/\sqrt{E\text{ [MeV]}}$. We assume that the TAO detector response is perfectly Gaussian.

For a given experimental configuration, specified by the near and far baselines ($L_\text{near}$ and $L_\text{far}$, respectively), we form the following $\chi^2$ function:
\begin{align}
    \chi^2 & = \sum_i^{N_\text{bins}} \left[ \left(\frac{M_i^N - \phi_i P_i^N\left(\sin^2 2\theta, \Delta m^2, \left\{ \eta_j \right\} \right)}{\sigma_i^N}\right)^2 \right. \nonumber \\
    & \left. \qquad + \left(\frac{M_i^F - \phi_i P_i^F\left(\sin^2 2\theta, \Delta m^2, \left\{ \eta_j \right\} \right)}{\sigma_i^F}\right)^2 \right] \nonumber \\
    & \qquad + \sum_j^{N_\text{sys}} \left( \frac{\eta_j}{\sigma_j} \right)^2
    \label{eq:chi2}
\end{align}
The components of this expression are as follows:
\begin{itemize}
    \item The $M_i^N$ and $M_i^F$ represent the numbers of pseudodata IBD counts in energy bin $i$ in the near and far detectors, respectively. These pseudodata are generated without oscillations.
    \item The $P_i^N$ and $P_i^F$ are the predicted numbers of events in energy bin $i$ for oscillation parameters \{$\sin^2 2\theta$, $\Delta m^2$\} and nuisance parameters $\{\eta_j\}$, which we discuss more below.
    \item The $\sigma_i^N$ and $\sigma_i^F$ are the statistical uncertainties associated with the predictions. We account for the nonzero background in forming $\sigma_i^N$ and $\sigma_i^F$; we write
    \begin{equation}
        \sigma_i^{N,F} = \sqrt{P_i^{N,F} + \frac{1+r}{r} B^{N,F}},
        \label{eq:def_sigma}
    \end{equation}
    where $B^N$ ($B^F$) is the number of background events per energy bin in the near (far) detector. The factor of $r$ in Eq.~\eqref{eq:def_sigma} is the ratio of reactor-off to reactor-on times; this arises from background subtraction. The expected raw number of events in each bin during the reactor-on period is $P_i^{N,F} + B^{N,F}$; during the reactor-off period, this is $rB^{N,F}$. Reweighting the (Poissonian) uncertainty on the latter by $r^{-1}$ and adding this in quadrature with the uncertainty on the former leads to Eq.~\eqref{eq:def_sigma}.
    We have simplified our analysis by taking $r=1$, i.e., that the reactor-on and reactor-off times are equal.
    \item The $\phi_i$ are nuisance parameters for the flux in each energy bin. Our analysis is \emph{flux-free} in that we do not introduce a prior on these nuisance parameters; while we use the Huber-Mueller (HM) flux model \cite{Mueller:2011nm, Huber:2011wv} to generate our pseudodata, our analysis is not sensitive to this choice. These unconstrained nuisance parameters reduce the number of degrees of freedom in the fit, leading to lower values of the $\chi^2$ and, ultimately, less statistical significance. However, sensitivities derived in this fashion are more robust than those that rely on a given flux model. We may analytically minimize the $\chi^2$ with respect to these nuisance parameters to obtain
    \begin{equation}
        \phi_i = \frac{ \dfrac{M_i^N P_i^N}{(\sigma_i^N)^2} + \dfrac{M_i^F P_i^F}{(\sigma_i^F)^2} }{\left(\dfrac{ P_i^N}{\sigma_i^N}\right)^2 + \left(\dfrac{ P_i^F}{\sigma_i^F}\right)^2 }.
    \end{equation}
    \item Lastly, the $\{\eta_j\}$ are the nuisance parameters describing our systematic uncertainties, which are assumed to be consistent with the systematics described by the STEREO Collaboration \cite{STEREO:2019ztb}:
    \begin{itemize}
        \item A 1.2\% uncorrelated normalization uncertainty.
        \item A 1.0\% uncorrelated energy scale uncertainty.
        \item A 0.3\% correlated energy scale uncertainty.
    \end{itemize}
    We expect this to characterize the capabilities of next-generation reactor experiments. We ignore the possibility of (uncorrelated) systematic uncertainties in the measured spectral shape in our initial studies; we discuss this further in Sec.~\ref{sec:exposure}.
\end{itemize}

We only consider simple $\Delta\chi^2$ statistics here. In recent years, it has been repeatedly emphasized that searches for sterile neutrino oscillations do not satisfy Wilks' theorem: simple $\Delta \chi^2$ statistics do not provide the proper coverage in the $\sin^2 2\theta$--$\Delta m^2$ parameter space, so converting $\Delta \chi^2$ values to confidence levels using Gaussian statistics is formally incorrect (see, e.g., Refs.~\cite{Agostini:2019jup,Coloma:2020ajw}). Consequently, we will avoid rigorously assigning confidence levels. That said, statistical methods that aim to address these deficiencies (e.g., the CL$_s$ method \cite{Qian:2014nha} and Feldman-Cousins method \cite{Feldman:1997qc}) are computationally intensive; performing these at scale is beyond the scope of this work. Therefore, we report our sensitivities as contour(s) along which $\Delta \chi^2 = 11.83$ --- for two degrees of freedom, this would correspond to $3\sigma$ if Wilks' theorem applied. In most cases, the true significance will be less than this.

\subsection{Research reactor optimization}
\label{sec:rr}

\begin{figure*}
\begin{subfigure}[t]{0.45\textwidth}
    \centering
    \includegraphics[width=\columnwidth]{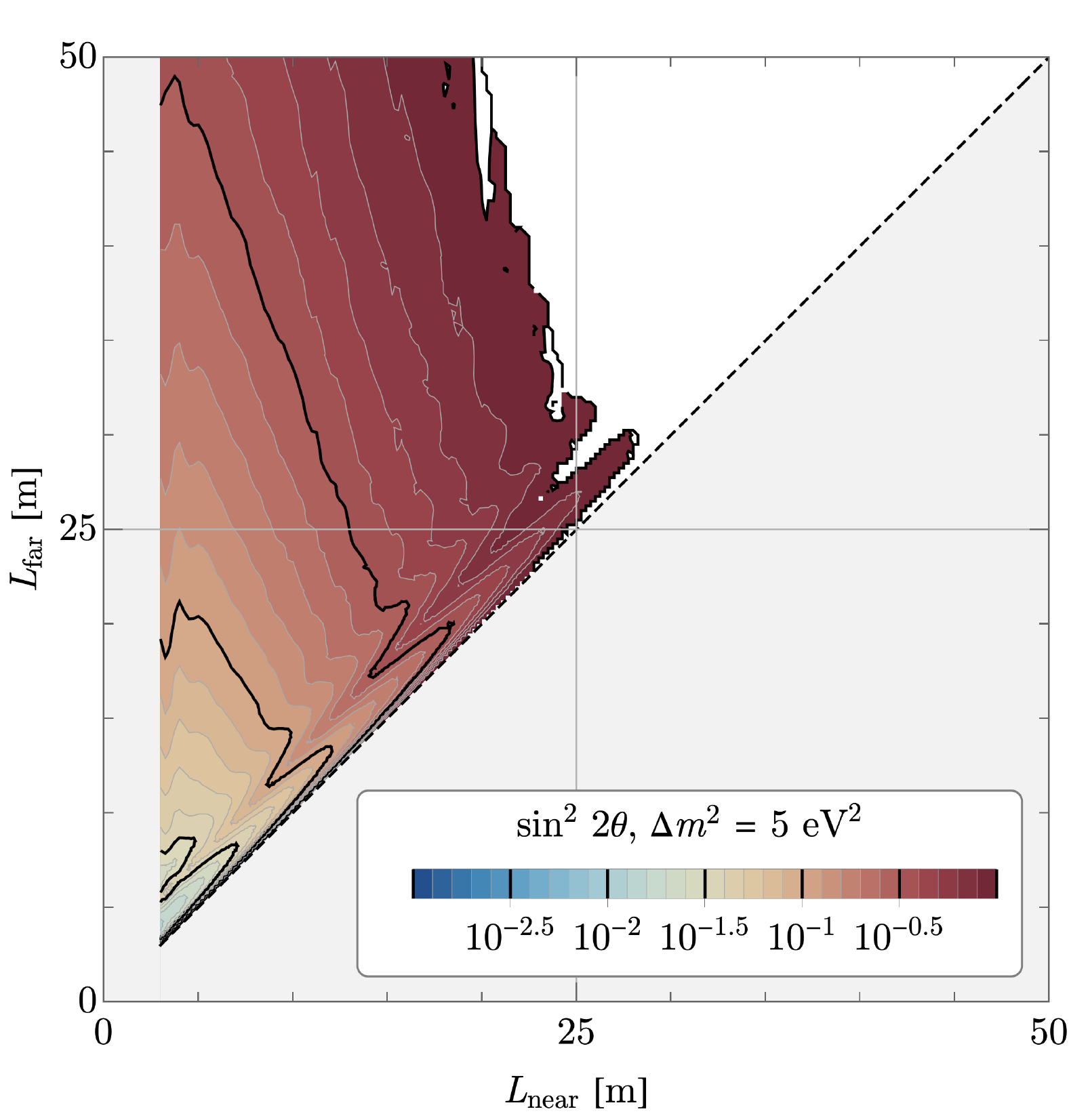}
    \caption{}
    \label{fig:prospect_research}
\end{subfigure}
\hspace{5mm}
\begin{subfigure}[t]{0.45\textwidth}
\centering
    \includegraphics[width=\columnwidth]{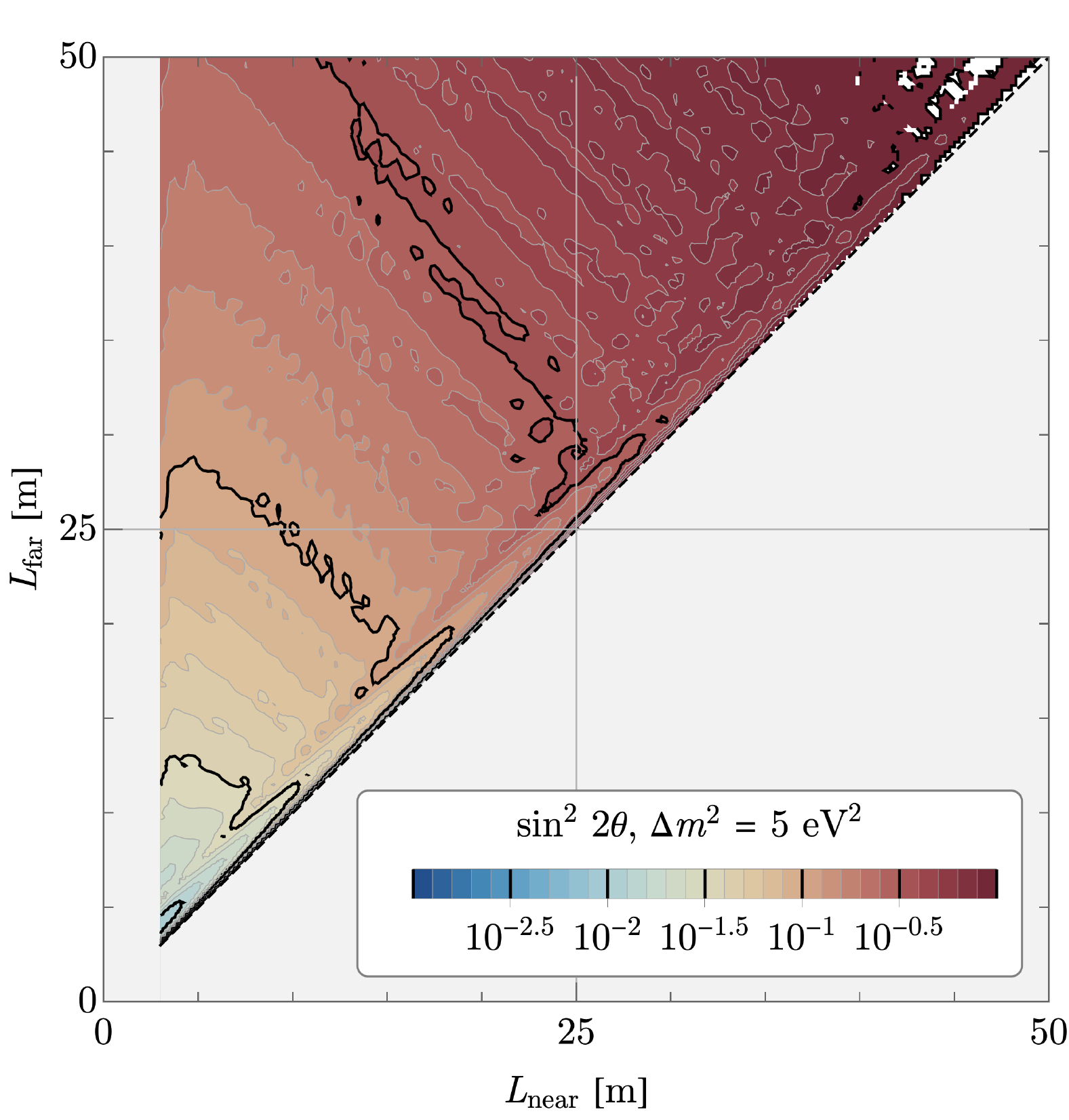}
    \caption{}
    \label{fig:tao_research}
\end{subfigure}
\caption{The sensitivity ($\Delta \chi^2 = 11.83$) of a hypothetical two-baseline research reactor experiment to oscillations with $\Delta m^2 = 5$ eV$^2$ as a function of its near and far baselines, $L_\text{near}$ and $L_\text{far}$. Panel (a) shows results for PROSPECT-like response, while panel (b) shows results for TAO-like response.}
\end{figure*}

We begin with research reactors. Because of their smaller size and because they are not attached to the heat-extraction apparatus of a commercial core, one could operate an experiment with much shorter baselines at a research core. For instance, the MINER experiment \cite{Agnolet:2016zir} can to operate within 1-3 m of a low-power (1 MW$_\text{th}$) reactor. However, it is more typical of the current generation of short-baseline experiments to operate in the range 6-10 m. We consider baselines as close as 3 m and as far as 50 m from a core whose radius is taken to be 20 cm and whose power is set to 100 MW$_\text{th}$. The total exposure is fixed at 1 yr, which we assume corresponds to 90 d of effective reactor-on time, accounting for a finite detector efficiency ($\sim50\%$). Moreover, because research reactors use high-enriched uranium fuel, the flux of antineutrinos is entirely attributable to $^{235}$U fission. We calculate the total number of events in each detector to be
\begin{align}
    N_\text{total}^\text{research} \approx \, & 1.5 \times 10^{6} \times \\
    & \left( \frac{\text{10 m}}{L} \right)^2 \left( \frac{m_\text{det} T \epsilon}{\text{5 ton yr}} \right) \left( \frac{P}{\text{100 MW}_{\text{th}}} \right), \nonumber
\end{align}
where $m_\text{det}$ is the detector mass, $T$ the total exposure time and $\epsilon$ an efficiency factor, which accounts for detector efficiency and reactor up time.

The downside of operating so close to the core is backgrounds. On one hand, operating close to the core subjects the detector to reactor-correlated neutron and gamma backgrounds. On the other, the detector is necessarily close to the surface  subjecting the detector to cosmogenic backgrounds. It is possible to mitigate either source of background, but this presents a fundamental challenge to the operation of such an experiment. We will assume that a background rate of 250 events per ton-day of reactor-on time spread uniformly over the analysis region at both detector sites. This is broadly consistent with the background rate achieved at PROSPECT \cite{Andriamirado:2020erz}.

In Figs.~\ref{fig:prospect_research} and \ref{fig:tao_research}, we show representative sensitivities to oscillations with $\Delta m^2 = 5$ eV$^2$ for a PROSPECT-like and TAO-like detector response, respectively. The color conveys the sensitivity ($\Delta \chi^2 = 11.83$) to $\sin^2 2\theta$ for each pair of near and far baselines; white space indicates baseline pairs for which there is no sensitivity. The central conclusion is that getting as close as possible to the reactor core is the key factor in improving the sensitivity. Indeed, the sensitivity is maximized when the spatial separation of the two detectors corresponds to half the oscillation wavelength $L_\text{osc}$ at $E_\nu \approx 4$ MeV, where the event rate is maximized:
\begin{equation}
    \label{eq:osc_wavelength}
    L_\text{osc} \equiv \frac{4 E_\nu}{\pi \Delta m^2} \approx \text{(2 m)} \times \left( \frac{E_\nu}{\text{4 MeV}} \right) \left( \frac{\text{5 eV}^2}{\Delta m^2} \right).
\end{equation}
Note that even in the case of a detector with TAO-like energy resolution, the sensitivity barely reaches $\sin^2 2\theta \sim 10^{-2}$ for the closest conceivable baseline of 3 m. Constructing an experiment this close to a core will be challenging; a less difficult case is a near baseline of $\sim 6$ m. In this case, the optimal sensitivity worsens slightly to $\sin^2 2\theta \sim 3 \times 10^{-2}$.

\subsection{Commercial reactor optimization}
\label{sec:cr}

\begin{figure*}
\begin{subfigure}[t]{0.45\textwidth}
    \centering
    \includegraphics[width=\columnwidth]{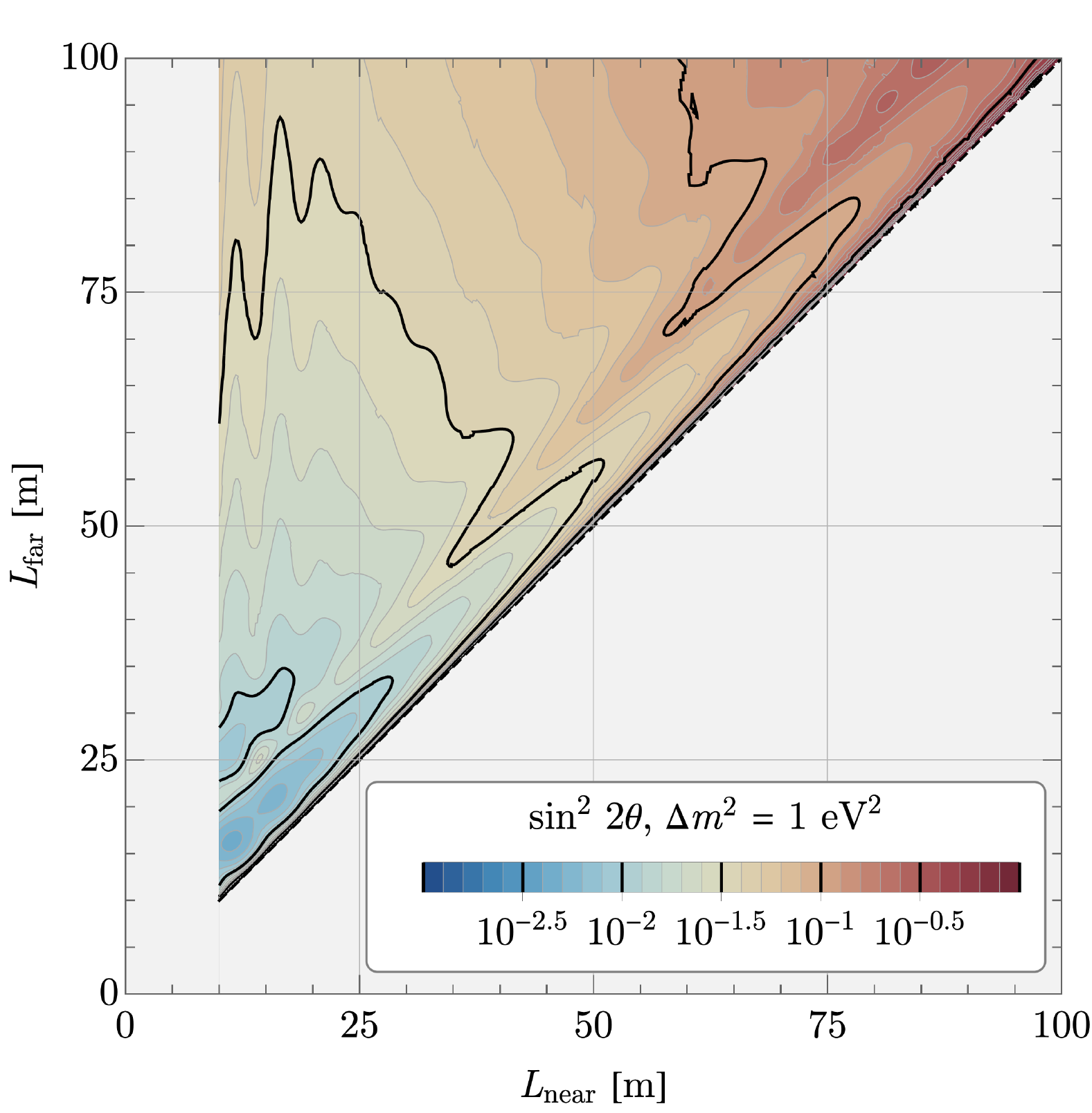}
    \caption{}
    \label{fig:prospect_commerical}
\end{subfigure}
\hspace{5mm}
\begin{subfigure}[t]{0.45\textwidth}
    \centering
    \includegraphics[width=\columnwidth]{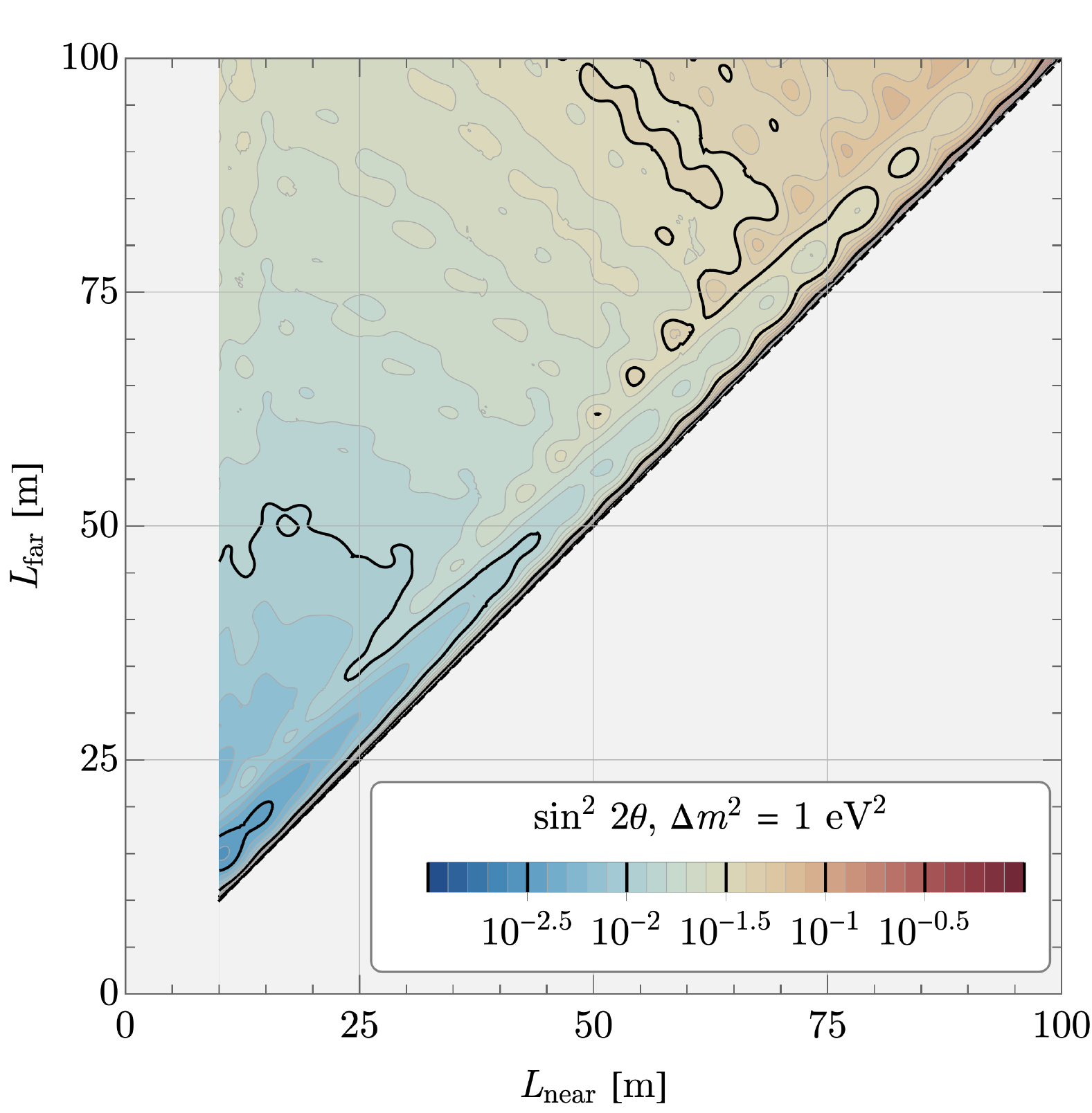}
    \caption{}
    \label{fig:tao_commercial}
\end{subfigure}
\caption{The sensitivity ($\Delta \chi^2 = 11.83$) of a hypothetical two-baseline commercial reactor experiment to oscillations with $\Delta m^2 = 1$ eV$^2$ as a function of its near and far baselines, $L_\text{near}$ and $L_\text{far}$. Panel (a) shows results for PROSPECT-like response, while panel (b) shows results for TAO-like response.}
\end{figure*}

Commercial reactors can be several orders of magnitude more powerful than research reactors, meaning that one does not need to operate an experiment at as short a baseline to have an appreciable event rate. This is fortunate, since one \emph{cannot} conduct an experiment at such close baselines, given the complicated layout of commercial reactor plants. We take the radius of the core to be 2.0 m, the core power to be 4.5 GW$_\text{th}$ and the exposure to be 1 yr, which we assume corresponds to 365 d of reactor-on time. Moreover, the fuel at commercial reactors is typically low-enriched uranium, meaning the antineutrino flux is a nontrivial combination of the fluxes from the four main fissile isotopes ($^{235}$U, $^{238}$U, $^{239}$Pu, $^{241}$Pu). We take the following values for the effective isotopic fission fractions:
\begin{equation*}
 \left\{^{235}\text{U},\,^{238}\text{U},\,^{239}\text{Pu},\,^{241}\text{Pu}\right\}=\left\{0.56,\,0.07,\,0.31,\,0.06 \right\}.
\end{equation*}
We calculate the total number of events in each detector to be
\begin{align}
    N_\text{total}^\text{commercial} \approx \, & 1.9 \times 10^{7} \times \\
    & \left( \frac{\text{20 m}}{L} \right)^2 \left( \frac{m_\text{det} T \epsilon}{\text{5 ton yr}} \right) \left( \frac{P}{\text{4.5 GW}_{\text{th}}} \right). \nonumber
\end{align}

The closest baseline that any experiment has attained at a commercial reactor is $\sim 10$ m at DANSS experiment \cite{Danilov:2019aef}, the result of the specific construction of the Kalinin Nuclear Power Plant. A more typical scenario is akin to the NEOS experiment \cite{NEOS:2016wee}, which sits $\sim24$ m from a commercial core in a tendon gallery at the Hanbit Nuclear Power Plant. We consider baselines between 10-100 m in our simulations. The impact of this lack of closeness on sterile neutrino searches is twofold. Firstly, because shorter baselines allow for searches for larger oscillation frequencies, the sensitivity of commercial reactor experiments to larger values of $\Delta m^2$ will be muted. Also, the larger core size also causes high-frequency oscillations to average out.  Secondly, because the experiment is further from the core, the backgrounds are less severe: reactor-correlated neutron and gamma backgrounds are substantially attenuated and higher overburdens can be achieved to reduce cosmogenic backgrounds. We assume a background rate of 90 events per ton-day of reactor-on time, roughly corresponding to the background rate measured at NEOS \cite{NEOS:2016wee}.\footnote{NEOS was able to achieve this background rate with an overburdern of $\sim20$ m.w.e. Generically, a detector at such short distances would require a decent amount of shielding in order to keep background rates low, but given that this has been achieved at NEOS, we assume it will be possible in future experiments, too. Given the large signal rates, we do not expect that our results would be appreciably different if the background rate were modestly larger than this.}

In Figs.~\ref{fig:prospect_commerical} and \ref{fig:tao_commercial}, we show sensitivities to oscillations with $\Delta m^2 = 1$ eV$^2$. The color scale is the same as in Figs.~\ref{fig:prospect_research} and \ref{fig:tao_research}. The sensitivity is improved for commercial reactors over research reactors, owing to the much larger event rate: for a given common baseline, the factors of $\sim45$ in assumed power and $\sim4$ in effective run time lead to a factor $\sim100-200$ in signal event rate at a commercial core relative to a research core. As before, the key factor in determining the sensitivity is allowing for as short a baseline as achievable, subject to the constraints of the facility. Moreover, we again observe that the separation of the detectors should be roughly half the oscillation wavelength in Eq.~\eqref{eq:osc_wavelength}. The best sensitivity shown for the TAO-like response corresponds to $\sin^2 2\theta \sim 3 \times 10^{-3}$, but even restricting the baselines to be no less than 25 m or using the PROSPECT response model (or both) yields a maximum sensitivity better than $\sin^2 2\theta \sim 10^{-2}$ for this value of $\Delta m^2$.

\subsection{Comparisons}
\label{sec:comp}
  
\begin{figure*}[t]
    \includegraphics[width=\linewidth]{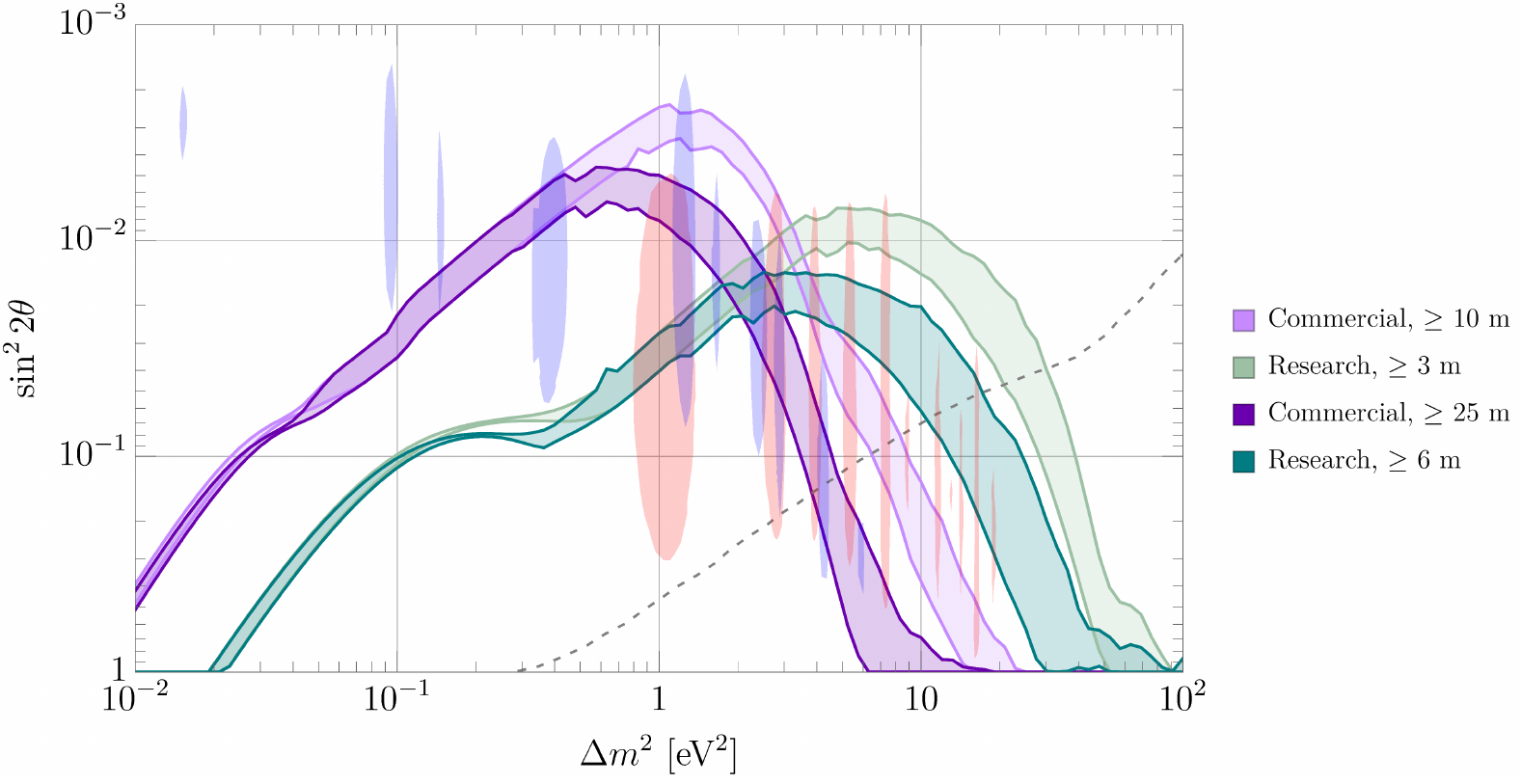}
    \caption{The baseline-optimized sensitivity ($\Delta \chi^2 = 11.83$) to $\sin^2 2\theta$ as a function of $\Delta m^2$. The purple (green) curves are for a commercial (research) reactor. The lighter band of either color represents the sensitivity with optimistic assumptions about the possible closest baseline; the darker bands represent more realistic assumptions. The weakest (strongest) sensitivity of each color corresponds to a PROSPECT-like (TAO-like) energy resolution. Also shown is the 95\% C.L. sensitivity from KATRIN (dashed gray), as well as the $3\sigma$-preferred regions from Neutrino-4 \cite{Neutrino4talk} (red) and from the global analysis of Ref.~\cite{Berryman:2020agd} (blue).}
    \label{fig:aggregate}
\end{figure*}

We aggregate our results for both the commercial and research reactor cases in Fig.~\ref{fig:aggregate}. The figure shows the baseline-optimized sensitivity ($\Delta \chi^2=11.83$) to $\sin^2 2\theta$ particular to a given $\Delta m^2$. We note that there is no achievable configuration that maximizes the sensitivity at every such value. The purple curves correspond to commercial reactors, whereas green curves are for research reactors. The lighter shade of either color exploits the full range of baselines, i.e., that a detector can be placed as close as possible to a given type of core (10 m for commercial, 3 m for research). The darker shade, in contrast, truncates the nearest baseline at what we believe is a more realistic distance (25 m for commercial, 6 m for research). The shading indicates the effect of the detector resolution, ranging between the PROSPECT (lower) and TAO (upper) responses.

For a fixed exposure, the sensitivity of a commercial reactor experiment exceeds that of a research reactor experiment by some $\mathcal{O}(1-10)$ factor, a consequence of the increased statistics at commercial experiments. However these experiments are most sensitive to oscillations with different values of $\Delta m^2$, a consequence of the different core sizes and permissible baselines. If one were optimizing a sterile neutrino search for $\Delta m^2 \lesssim 2-3$ eV$^2$, then one would ultimately elect to operate the experiment at a commercial reactor; conversely, for larger values of $\Delta m^2$, one would choose a research reactor. This figure concretely demonstrates the benefit of putting the detector closer to the core for probing high-frequency oscillations --- a transition occurs around $\Delta m^2 \sim 0.5$ eV$^2$ ($\sim 2$ eV$^2$) for commercial (research) reactors where the advantage of higher statistics and shorter baselines dissipates.

To contextualize these sensitivities, we also show the projected final 95\% C.L. sensitivity of the ongoing KATRIN tritium-decay experiment \cite{KATRIN:2020dpx} in dashed gray. While a research core would be the preferred method up to $\Delta m^2 \sim 30$ eV$^2$ if one could realize ultrashort baselines, KATRIN is ostensibly the preferred technique for searches above $\Delta m^2 \sim 15$ eV$^2$ for more realistic configurations. However, there is an important caveat in interpreting the KATRIN sensitivity: this depends on the prior on the effective neutrino mass squared, $m_\nu^2$, that one extracts from the experiment. If one insists that $m_\nu^2 \ge 0$ -- a sensible physical criterion -- then one derives the sensitivity shown in the figure. However, one may reasonably insist on a different prior on this quantity, given that historical tritium-decay experiments have often found negative best-fit values of this parameter; see Ref.~\cite{Formaggio:2021nfz} and references therein. If one were to allow $m_\nu^2$ to take any value -- positive or negative -- to avoid a biased measurement, then the resulting sensitivity changes; this is depicted in Fig.~3 of Ref.~\cite{KATRIN:2020dpx} for the exclusion based on current data. We are unaware of any study of this effect on the ultimate sensitivity of KATRIN and do not attempt one here, but this suggests that the constraint may not be robust below $\Delta m^2 \sim 30$ eV$^2$. If this is the case, then research reactor experiments may be the best path forward to exploring this portion of the parameter space.

The last two components of Fig.~\ref{fig:aggregate} are the regions of parameter space preferred by analyses of separate existing datasets. The first is the $3\sigma$-preferred region from an analysis of the Neutrino-4 experiment \cite{Neutrino4talk}, shown in red shading. The second is the $3\sigma$-preferred\footnote{What this reference calls ``$3\sigma$'' should more appropriately be called the ``$\Delta \chi^2 = 11.83$'' exclusion.} region from a recent global analysis of Bugey-3, DANSS, Daya Bay, Double Chooz, NEOS and RENO from Ref.~\cite{Berryman:2020agd}, shown in blue shading. These are the regions that next-generation reactor experiments will target --- indeed, the presence of closed $3\sigma$ contours in existing analyses is what inspires the next generation of experiments in the first place.\footnote{There are, of course, practical applications of antineutrino physics that will require research and development; see, e.g., Ref.~\cite{Bergevin:2019tcg} and references therein. However, these technologies depend on a precise understanding of isotopic antineutrino fluxes and are thus inextricably coupled to the oscillation anomalies.} Our results suggest that the optimal strategy to probe these regions would be to build a commercial reactor experiment to probe below $\Delta m^2 \sim 2$ eV$^2$ and a research reactor experiment to probe above this, and that they should both be as close as can be achieved to their respective cores.

We have considered the effect of bin size on Fig.~\ref{fig:aggregate}. Specifically, we have looked at bins of width 100 keV for both a PROSPECT-like and TAO-like response model at both a commercial and research reactor facility. On balance, the changes induced are not markedly different from Fig.~\ref{fig:aggregate}. We have also investigated the effect of 30-keV bins on analyses with our TAO-like response and again find no appreciable shift in sensitivity.

\subsection{Exposure and systematics limitation}
\label{sec:exposure}

One can ask how the sensitivities shown in Fig.~\ref{fig:aggregate} can be improved by either operating the experiment for a longer period of time or by constructing a more massive detector. To study this, we must include an effect that we have ignored to this point. While our flux-free analysis does not introduce any prior uncertainties on the flux, one should account for systematic uncertainties on the measured spectral shape, in addition to the systematics that we have already included. We assume that this systematic is uncorrelated between energy bins and that next-generation experiments can achieve shape uncertainties as low as 0.5\%, consistent with the value claimed by TAO \cite{Abusleme:2020bzt}. This is introduced to our $\chi^2$ in Eq.~\eqref{eq:chi2} via the replacement
\begin{equation}
    \left(\sigma^{N,F}_i\right)^2 \to \left(\sigma^{N,F}_i\right)^2 + (P_i^{N,F} \sigma_\text{sh})^2,
\end{equation}
where $\sigma_\text{sh}$ represents the shape systematic.

\begin{figure}[t!]
    \includegraphics[width=\columnwidth]{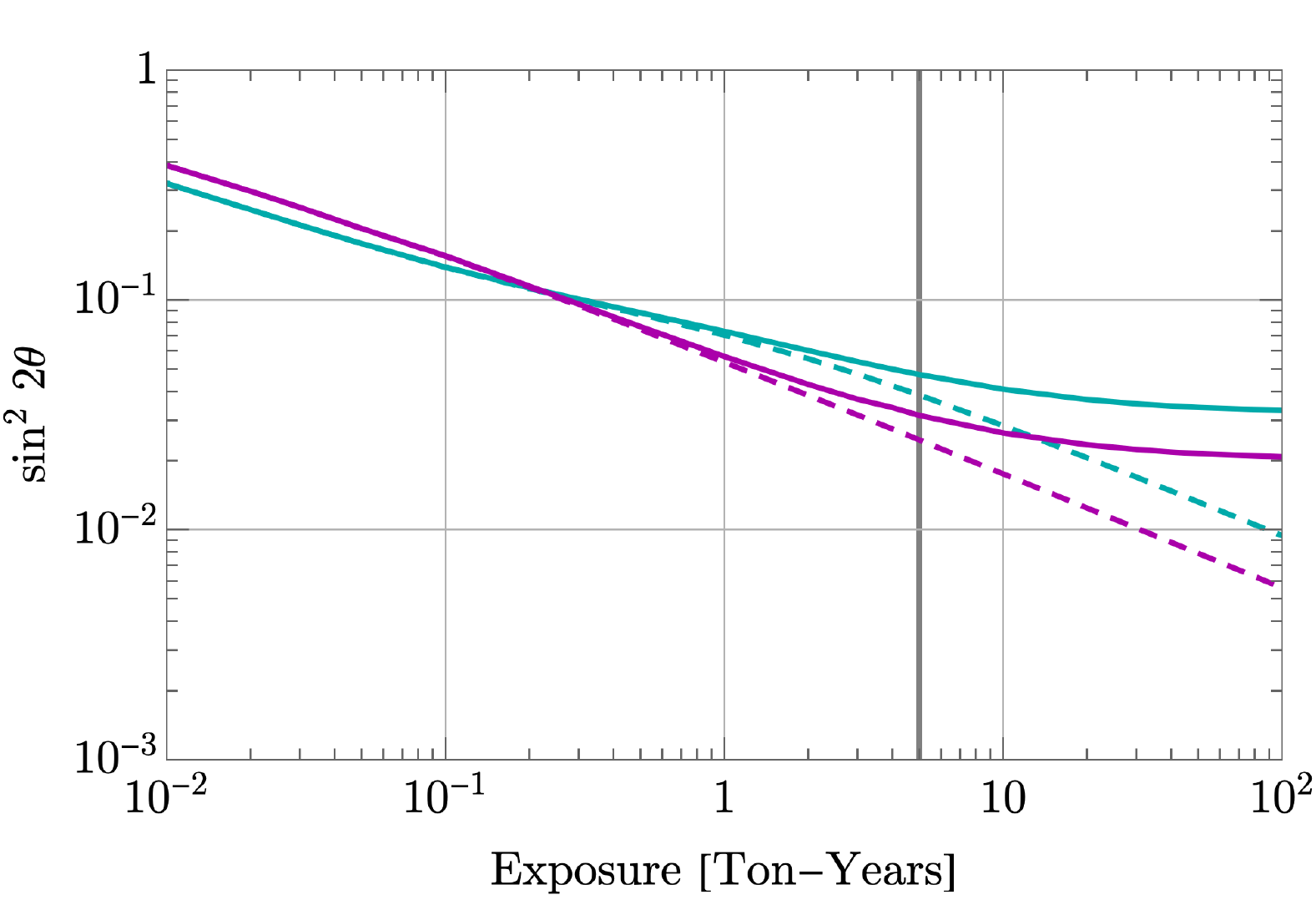}
    \caption{The sensitivity ($\Delta \chi^2 = 11.83$) to $\sin^2 2\theta$ as a function of exposure for a research reactor experiment with $L_\text{near} = 6$ m and $L_\text{far} = 9$ m. Cyan (magenta) curves correspond to $\Delta m^2 = 1$ eV$^2$ ($\Delta m^2 = $ 5 eV$^2$). Dashed curves correspond to vanishing shape systematic uncertainty; solid curves take this to be 0.5\%. The vertical, gray line indicates our benchmark 5-ton-years exposure, at which value each detector observes $3.2\times10^{5}$ events.}
    \label{fig:exposureU}
\end{figure}

To demonstrate the interplay between the exposure and a finite shape systematic, we consider a pseudoexperiment at a research reactor with a near baseline of 6 m and a far baseline of 9 m, roughly corresponding to the near and far baselines at PROSPECT, that employs the TAO response model. In Fig.~\ref{fig:exposureU}, we show the sensitivity to $\sin^2 2\theta$ for fixed values of $\Delta m^2$ -- 1 eV$^2$ in cyan and 5 eV$^2$ in magenta -- as a function of exposure. The vertical, gray line corresponds to the 5-ton-years exposure used in our previous analyses. The dashed curves correspond to vanishing shape uncertainty. For these, as the exposure is increased, the sensitivity improves without bound: the increase in statistics leads to an increasingly precise determination of the event rate at each position and in each energy bin, resulting in sensitivity that scales as $\sim \left(\text{exposure}\right)^{-1/2}$. In contrast, the solid curves show the sensitivity including a 0.5\% shape uncertainty. As the number of raw counts is increased, the fractional statistical uncertainty eventually becomes eclipsed by the shape uncertainty -- the experiment becomes systematics limited. Past this point, the sensitivity saturates at some finite value. Inspection of Fig.~\ref{fig:exposureU} reveals that for our nominal 5-ton-year exposure, the experiment is already transitioning from being statistically limited to being systematically limited. Consequently, increasing the exposure beyond this value results in a negligible increase in sensitivity to $\sin^2 2\theta$.

\section{Sterile Neutrino Sensitivity at TAO}
\label{sec:tao}
\setcounter{equation}{0}

TAO is a future experiment whose main purpose is to produce a model-independent reference spectrum for the physics program at JUNO~\cite{An:2015jdp}. It is planned to start taking data in 2022 and is expected to achieve an unprecedented experimental resolution of the order $\sim1.0\%/\sqrt{E\text{ [MeV]}}$. Though it is a \emph{single-volume} detector, its fiducial volume may be \emph{virtually} segmented, allowing for a multiple-baseline measurement of the sort outlined in the previous section. The segmentation is possible due to improvements in the coverage and efficiency of the photo sensors, as well as the precise timing resolution; see, e.g., Sec.~8.5 of Ref.~\cite{Abusleme:2020bzt}. This is a novel capability that has not been employed by existing searches for sterile neutrino. We have previously considered two detectors located at arbitrary baselines; here, on the other hand, we consider a fixed (core-to-detector) baseline position of 30 m and compare the spectra measured in each of its virtual segments. We will virtually segment the detector into either two or four equal-volume segments; the displacements between the barycenters and the center of the detector are ($\pm 3/8$ $R$) and ($\pm$ 0.170 $R$, $\pm$ 0.580 $R$), respectively, $R = 65$ cm being the radius of the fiducial volume at TAO. While virtual segmentation is an impressive capability of this detector, the fixed geometry is a nontrivial restriction on a multiple-baseline measurement.

We assume a total of 5-ton-year exposure at a commercial core with a 2.0-m radius and a power of 4.5 GW$_{\text{th}}$, in broad agreement with the TAO conceptual design report (CDR) \cite{Abusleme:2020bzt}. The fuel fractions of the core are assumed to be the same as our green-field studies (Sec.~\ref{sec:gf}). Moreover, a total of 60 bins between 2.0 to 8.0 MeV in prompt energy are considered. Background events are spread uniformly over the energy bins, assuming a total of 90 events per ton-day, consistent with those at NEOS \cite{NEOS:2016wee}; we do not consider shape uncertainties on the background. Systematic uncertainties in this experiment are assumed to be consistent with those at STEREO \cite{STEREO:2019ztb}. Furthermore, a 0.5\% bin-to-bin uncorrelated shape uncertainty is taken into account.

For our experimental configuration, we form the following $\chi^2$ function:
\begin{align}
    \chi^2 & = \sum_{i}^{N_\text{bin}} \sum_{j}^{N_\text{seg}} \left(\frac{M_{i,j} - \phi_i P_{i,j}\left(\sin^2 2\theta, \Delta m^2, \left\{ \eta_\ell \right\} \right)}{\sigma_{i,j}}\right)^2 \nonumber \\
    & \qquad + \sum_\ell^{N_\text{sys}} \left( \frac{\eta_\ell}{\sigma_\ell} \right)^2 .
    \label{eq:chi2_tao}
\end{align}
This expression closely resembles Eq.~\eqref{eq:chi2}, the salient difference being that we now index the $N_\text{seg}$ ($=2,\,4$) segments with $j$. We account for the nonzero background and shape uncertainty in a similar fashion as Eq.~\eqref{eq:def_sigma} by defining $\sigma_{i,j}$ as 
    \begin{equation}
        \sigma_{i,j}^2 = P_{i,j}(1+\sigma_{\text{sh}}^2 \times P_{i,j}) + 2 B_{i,j},
        \label{eq:def_sigma_tao}
    \end{equation}
were $\sigma_{\text{sh}}$ is the bin-to-bin shape uncertainty and $B_{i,j}$ is the (uniform) background in each energy bin $i$ of segment $j$.\footnote{We are again assuming that the reactor-off and reactor-on times are equal, leading to the factor of 2 in Eq.~\eqref{eq:def_sigma_tao}} The $\phi_i$ are flux nuisance parameters in each energy bin. As before, we analytically minimize the $\chi^2$ with respect to these nuisance parameters to obtain
    \begin{equation}
        \phi_i = \frac{ \sum_{j} \dfrac{M_{i,j} P_{i,j}}{(\sigma_{i,j})^2}}{\sum_{j} \left(\dfrac{ P_{i,j}}{\sigma_{i,j}}\right)^2} ~,
        \label{eq:phi_tao}
    \end{equation} 
where $j$ runs over the detector segments. Lastly, the $\{\eta_\ell\}$ are the nuisance parameters describing our systematic uncertainties.

In Ref.~\cite{Abusleme:2020bzt}, the HM flux model \cite{Mueller:2011nm,Huber:2011wv} is used to obtain the antineutrino energy spectrum with an introduction of a 5\% bin-to-bin uncorrelated shape uncertainty on the flux. While we, too, use the HM fluxes, our analysis is not sensitive to this choice; by using the $\phi_{i}$, our analysis employs a data-to-data spectral comparison, as compared to a data-to-model one. This removes any model dependence on our analysis, but reduces the number of degrees of freedom in the fit, leading to lower values of the $\chi^2$ and, ultimately, less inferred statistical significance. Therefore, we expect the sensitivities obtained in this study to be less aggressive but more robust. Another important difference between our analysis and that of Ref.~\cite{Abusleme:2020bzt} is the assessment of the backgrounds: the collaboration publishes a combined rate of accidental, fast neutron and $^9$Li/$^8$He decay backgrounds of $\sim$ 450 events per ton-day, a more conservative figure than that assumed in our study. We have consciously elected to be more optimistic, given that the 90 events per ton-day we consider has been achieved at NEOS \cite{NEOS:2016wee}. We also remind the reader that we only communicate sensitivities as contours of constant $\Delta \chi^2$, whereas Ref.~\cite{Abusleme:2020bzt} provides the results of Gaussian CL$_s$ analysis. This is not, strictly speaking, an apples-to-apples comparison of sensitivities; however, the gross features of the analyses are distinct enough that one can still draw important qualitative conclusions.

\begin{figure}[!t]
    \includegraphics[width=\columnwidth]{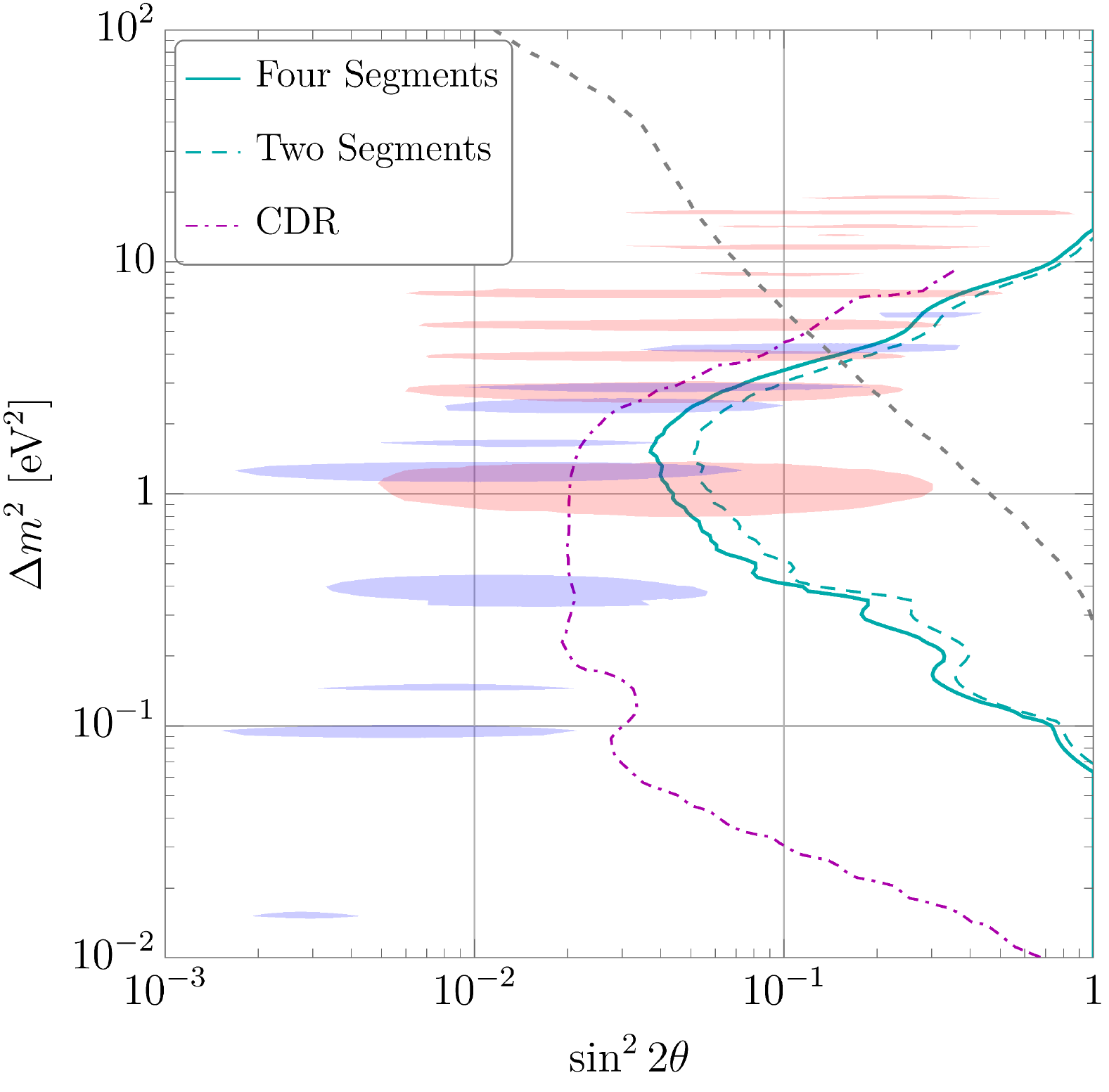}
    \caption{The sensitivity ($\Delta \chi^2 = 11.83$) to a sterile neutrino at TAO. Cyan curves correspond to our analysis with four (solid) or two (dashed) virtual segments. The dotted-dashed, magenta curve corresponds to the 99.7\% CL$_s$ sensitivity presented in Ref.~\cite{Abusleme:2020bzt}. Also shown are the 95\% C.L. sensitivity from KATRIN (dashed gray), as well as the $3\sigma$-preferred regions from Neutrino-4 \cite{Neutrino4talk} (red shading) and from Ref.~\cite{Berryman:2020agd} (blue shading).}
    \label{fig:TAO_segmented}
\end{figure}

We show our results in Fig.~\ref{fig:TAO_segmented}. The dashed and solid cyan curves show the sensitivity ($\Delta \chi^2 = 11.83$) to sterile neutrinos for our analyses with two and four segments, respectively. The dotted-dashed, magenta curve corresponds to the 99.7\% CL$_s$ sensitivity from the TAO CDR \cite{Abusleme:2020bzt}. We also show the 95\% C.L. sensitivity from KATRIN in dashed gray, as well as the $3\sigma$-preferred regions from Neutrino-4 \cite{Neutrino4talk} and from Ref.~\cite{Berryman:2020agd}, in red and blue shading contours, respectively.

In the region $\Delta m^2 \leq 1$ eV$^2$, our analysis is substantially less sensitive than that of the CDR. This is a consequence of the (in)dependence on the flux model in these analyses. By comparing directly to the HM flux model, the CDR analysis is sensitive to changes in the antineutrino spectrum induced over the $\sim30$-m baseline between the core and the detector, affording sensitivity to lower values of $\Delta m^2$. In contrast, by comparing data to data, our analysis is sensitive only to changes induced over the $\sim1$-m length scale spanned by the detector. Therefore, sensitivity to longer-wavelength oscillations is inevitably muted, and the barycenter of the sensitivity curve shifts to higher values of $\Delta m^2$. Furthermore, the addition of more segments improves the overall sensitivity; the gains, however, are modest. On one hand, increasing the number of segments increases the number of degrees of freedom in the fit. On the other, this also increases the (statistical) uncertainty of the contents of any given energy bin. The former evidently outweighs the latter, though only marginally. Lastly, we note that the results of this analysis are largely independent of the binning, despite the exceptional resolution available. The reason for this is clear: there exists a trade-off between increasing the number of bins and the statistical power of a given bin. For the nominal operation parameters we have employed, decreasing the bin size beyond 100 keV does not lead to a meaningful gain in sensitivity. Of course, increasing the exposure beyond the nominal 5 ton-years would lead to an improvement, but the gains in $\sin^2 2\theta$ are expected to be $\sim\mathcal{O}(1)$ before systematics become dominant.

On balance, the sensitivity to sterile neutrinos presented in this study is less optimistic than the TAO CDR, but this is a consequence of having fundamentally different search strategies. Flux model independence is central to our analysis, as it was in the previous section. We believe this is conceptually more robust, though we have made some assumptions that are more optimistic than those presented in the CDR.

\section{Lithium-8 doping of research reactors}

\label{sec:li8}
\setcounter{equation}{0}

As we have seen in the previous sections, the sensitivity to oscillations with $\Delta m^2 \gtrsim 2$ eV$^2$ is significantly improved if the near baseline is minimized as much as possible, due to (1) the increased event rate, and (2) high-frequency oscillations not averaging out. However, the physical layout of the facility severely constrains the possible location of a detector.
Notice that because oscillations depend on the quantity $\Delta m^2 L/E$, the effect of reducing the baseline can be emulated by increasing the energies of the interacting antineutrinos. However, the energy spectrum of the antineutrinos emitted in fission is, of course, immutable.\footnote{This is not strictly speaking true: the spectrum changes slightly as the composition of the reactor's fuel evolves. Here, we are referring to the impossibility of manipulating a given $\beta$ decay into yielding more energetic neutrinos.} In order to reach higher energies, then, one would need to consider a separate source of antineutrinos.

We consider precisely such a source, in the form of antineutrinos produced via the decays of $^8$Li, produced by neutron capture on $^7$Li. The main advantage is that the endpoint energy of these decays is $\sim13$ MeV; this provides the high energies desired to extend the sensitivity of a sterile neutrino search to higher $\Delta m^2$.
Using lithium at a reactor to probe neutrino oscillations was previously suggested in Ref.~\cite{Lyashuk:2016lpn}; we aim to robustly study the sensitivity to a sterile neutrino within this scheme.
A similar proposal exists in the form of IsoDAR \cite{Conrad:2013sqa,Alonso:2017fci}, but implementing a $^8$Li source at a nuclear reactor has the advantage that no new facilities are required to produce the neutrons that ultimately drive the source. If one can load the reactor core with enough $^7$Li, then the excess neutrons produced in the core can, in principle, be used to generate these $\beta$ decays. 

\begin{figure}[t]
    \includegraphics[width=\columnwidth]{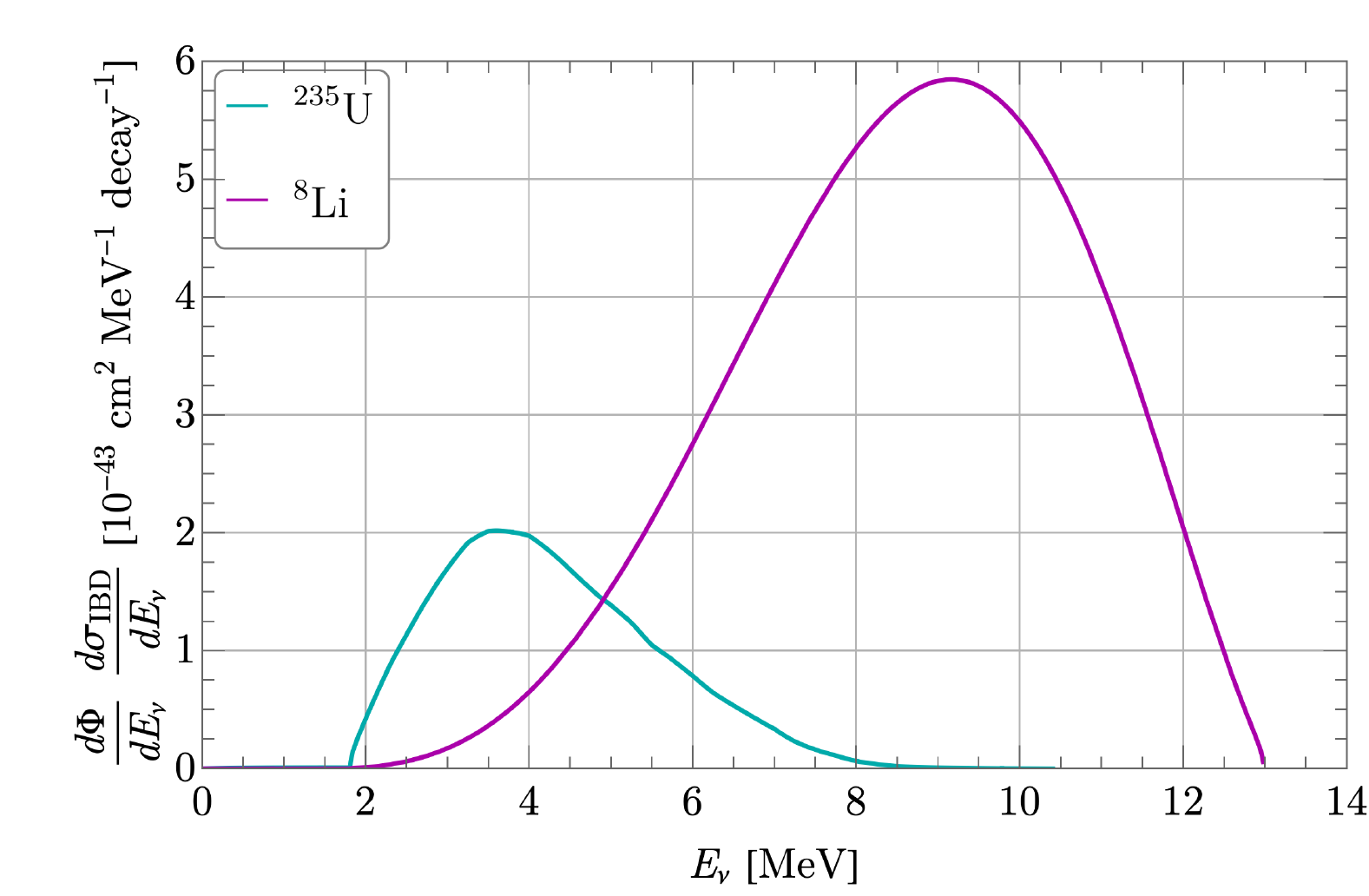}
    \caption{The flux-weighted cross section per decay for IBD detection of antineutrinos from $^{235}$U fission (cyan) and from $^8$Li $\beta$ decay (magenta). }
    \label{fig:lithium_spec}
\end{figure}

In Fig.~\ref{fig:lithium_spec}, we show the flux-weighted IBD cross section per $^8$Li decay against the same per $^{235}$U fission. Because the IBD cross section grows roughly quadratically in energy, the events associated with $^8$Li $\beta$ decay are disproportionately skewed to higher energies than those from fission. This results in a larger integrated flux-weighted cross section \emph{per decay}; fewer antineutrinos are emitted per decay, but they are much more inclined to interact with the detector. However, while the per-decay rate of $^8$Li events exceeds that of $^{235}$U, it is unlikely that the absolute rate of $^8$Li decay in an experiment could be arranged to be as large as the rate of $^{235}$U fission. More broadly, there are a number of challenges in implementing $^8$Li decays at nuclear reactors:
\begin{itemize}
    \item Inserting $^8$Li into a commercial core is essentially out of the question. This would require a significant interruption of operation of a core used to generate electricity and, consequently, income for the operator. 
    \item Research reactor cores, in contrast, are  constructed such that one may insert a sample of some material into the core to absorb some excess neutrons without disrupting the core. However, if one wanted to achieve lithium decay rates equal to the fission rate, then one would essentially need to absorb \emph{every} excess neutron over the operation of the experiment. Apart from reactor physics constraints, this would monopolize the facility, that is, no neutrons would be available for any other user.
    \item Lastly, while natural lithium is dominantly ($\sim92.4\%$) $^7$Li, the thermal neutron capture cross section for $^6$Li is roughly three orders of magnitude larger than that of $^7$Li. Therefore, one would need high-purity $^7$Li -- around 99.9\% -- to ensure that the majority of neutrons will ultimately produce $^8$Li; to ensure that 90\% of neutrons capture on $^7$Li, this increases to nearly 99.99\% purity. Scaling from previous results on Li loading of HFIR~\cite{conant} a 10\% fraction of neutrinos may be possible in this case. Note that high-purity $^7$Li is available in large quantities and considered as coolant in molten salt reactors.
\end{itemize}
In light of these restrictions, we expect that it may be possible to achieve a rate of one $^8$Li decay for every five to ten fission events at a research core.

We employ a pseudoexperiment with $L_\text{near} = 6$ m and $L_\text{far} = 9$ m. The total exposure of the experiment is fixed to be 5 ton-years, apportioned between the two detectors according to their inverse-squared baseline, and we continue to assume that one year at a research reactor corresponds to three months of reactor-on time with an average power of 100 MW$_\text{th}$. The rate of background events is still assumed to be 250 events per ton-day \emph{in the region 2.0-8.0 MeV prompt energy}, i.e., 41.7 counts per ton-day per MeV; we assume that this rate extends out to 13 MeV. We further assume a TAO-like energy resolution and employ 100-keV bins between 2.0 and 12.0 MeV in prompt energy to take full advantage of this resolution. The systematics budget is the same as in previous studies and we ignore any bin-uncorrelated shape uncertainties.

\begin{figure}[t]
    \includegraphics[width=\columnwidth]{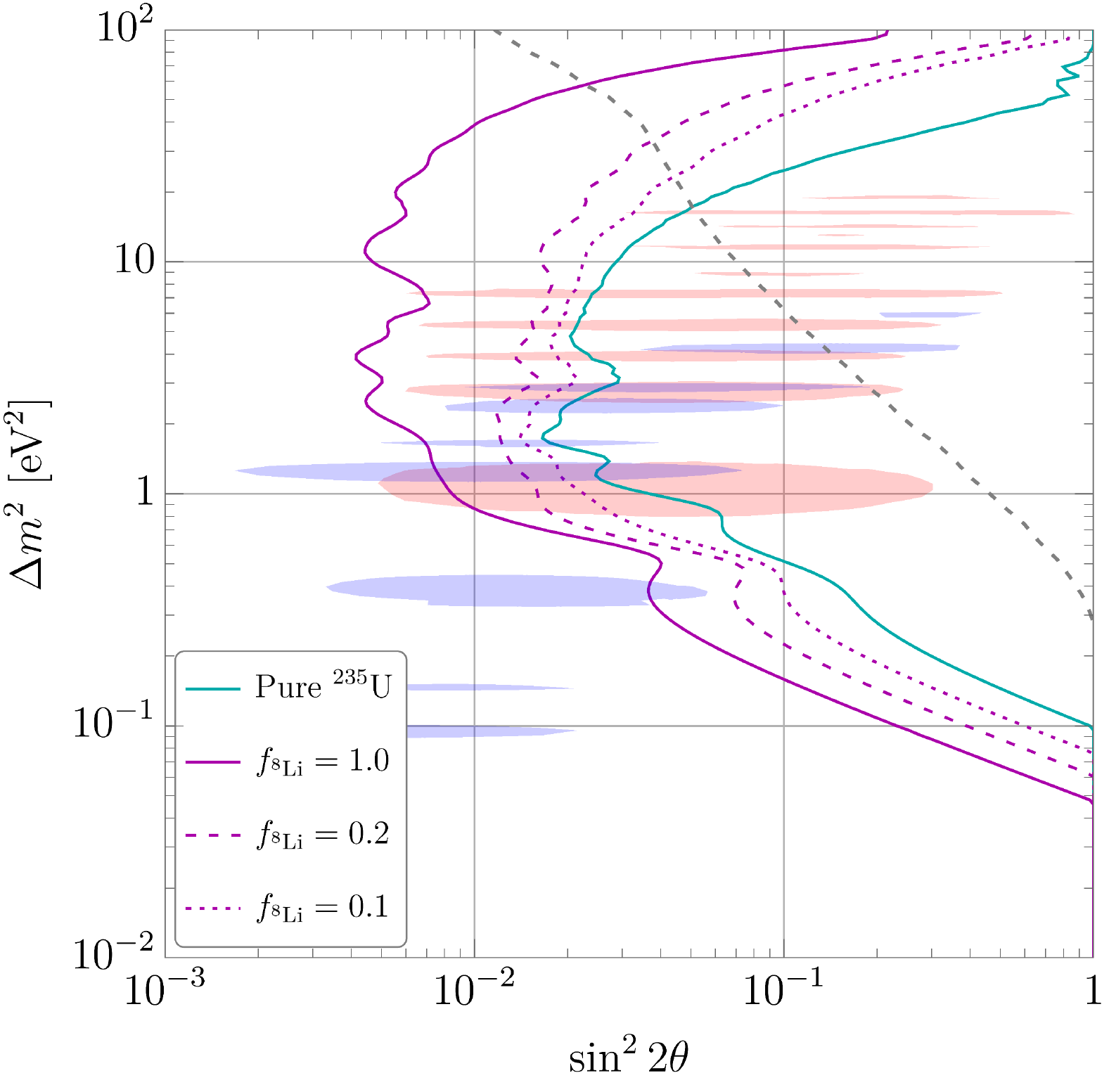}
    \caption{The sensitivity ($\Delta \chi^2 = 11.83$) of a lithium-loaded research reactor to a sterile neutrino. The sensitivity without $^8$Li is shown in cyan while the magenta curves assume varying numbers of lithium decays per fission, $f_{^8\text{Li}}$: 1.0 (solid), 0.2 (dashed) and 0.1 (dotted). We also show the 95\% C.L. sensitivity from KATRIN (dashed gray), as well as the $3\sigma$-preferred regions from Neutrino-4 \cite{Neutrino4talk} (red shading) and Ref.~\cite{Berryman:2020agd} (blue shading).}
    \label{fig:lithium}
\end{figure}

The results are shown in Fig.~\ref{fig:lithium}. The cyan curve shows the sensitivity ($\Delta \chi^2 = 11.83$) in the absence of any added lithium. We note that this curve is slightly more sensitive to oscillations with $\Delta m^2 \gtrsim 20$ eV$^2$ than the configuration-optimized sensitivity in Fig.~\ref{fig:aggregate} suggests, owing to the smaller bin size used here. Aside from the region around $\Delta m^2 \sim 3-4$ eV$^2$, this sensitivity is not too far from the optimal sensitivity for a 5-ton-year exposure; we expect this to be representative of the capabilities of a multiple-baseline experiment with baselines of this order. The magenta curves introduce varying numbers of $^8$Li decays per $^{235}$U fission, which we denote $f_{^8\text{Li}}$: solid corresponds to $f_{^8\text{Li}} = 1$, dashed to $f_{^8\text{Li}} = 0.2$ and dotted to $f_{^8\text{Li}} = 0.1$. We also show the projected sensitivity from KATRIN (dashed gray) and the $3\sigma$-preferred regions from Neutrino-4 (red shading) and from Ref.~\cite{Berryman:2020agd} (blue shading). The benefit of including lithium decays is clear: for $f_{^8\text{Li}} = 1$, the sensitivity improves by as much as an order of magnitude, particularly above $\Delta m^2 \gtrsim 5$ eV$^2$. In light of the aforementioned issues in the interpretation of the KATRIN constraint in the region $\Delta m^2 \lesssim 30$ eV$^2$, reactor experiments may be the only means by which to robustly probe a sterile neutrino in this region of parameter space --- and a highly lithium-loaded reactor may even exceed the sensitivity of KATRIN above this value.

\begin{figure}[t]
    \includegraphics[width=\columnwidth]{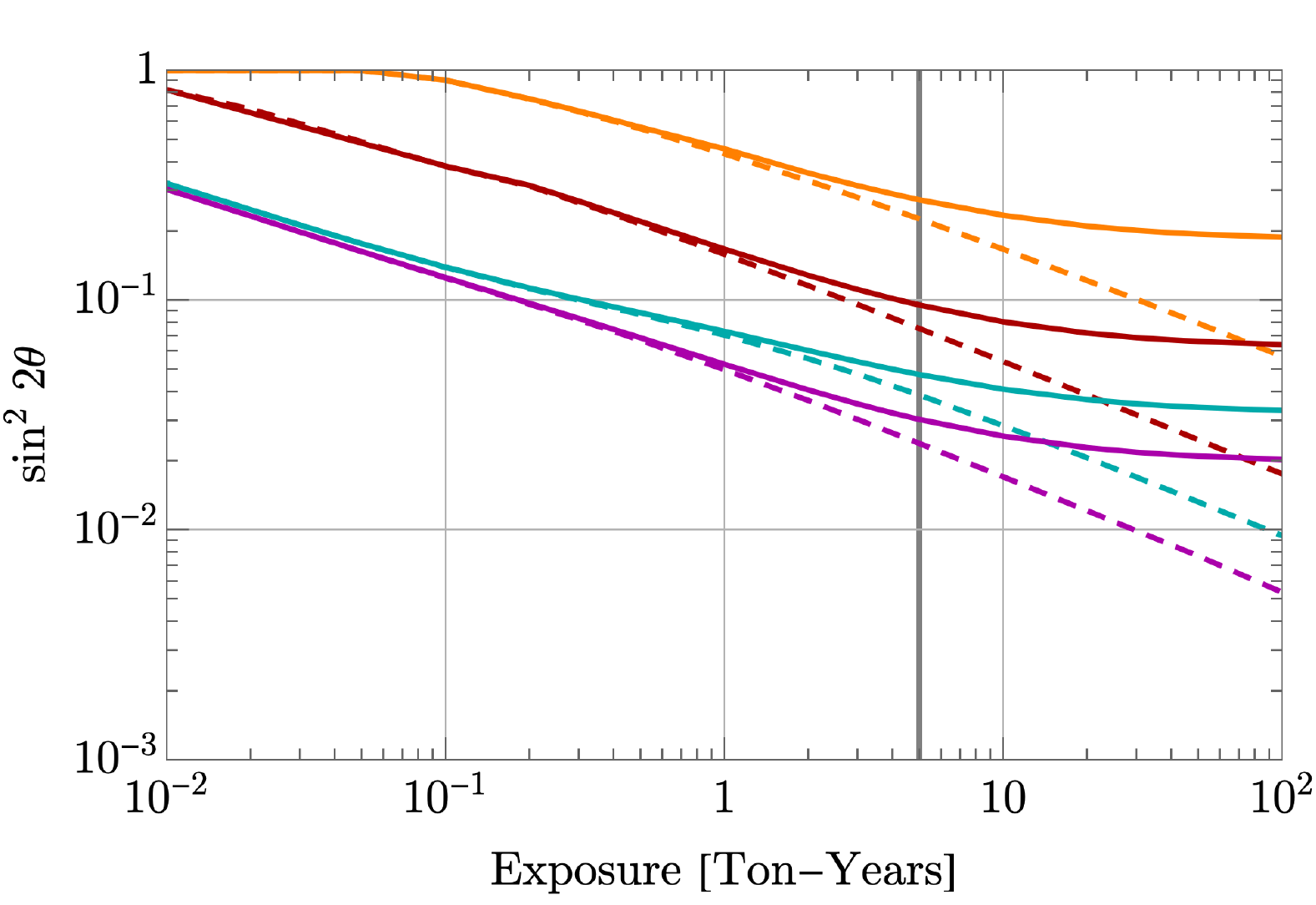}
    \caption{The evolution of the sensitivity ($\Delta \chi^2 = 11.83$) as a function of exposure for our lithium-enhanced research reactor pseudoexperiment. Cyan and orange curves assume only $^{235}$U is present for $\Delta m^2 = 1$ eV$^2$ and 30 eV$^2$, respectively; magenta and red curves introduce lithium with $f_{^8\text{Li}} = 0.1$. Dashed curves ignore shape systematics and solid curves assume a 0.5\% shape uncertainty. The vertical gray line indicates a 5-ton-year exposure.}
\label{fig:exposureLi}
\end{figure}

However, for reasons laid out above, there are a number of impediments to achieving $f_{^8\text{Li}} = 1$. It would be more realistic to expect $f_{^8\text{Li}}$ in the range $0.1-0.2$, in which case the dashed and dotted magenta lines should guide our interpretation of Fig.~\ref{fig:lithium}. The gains in sensitivity are more modest in this case. To compensate for this, one could envision increasing the exposure of the experiment. In Fig.~\ref{fig:exposureLi}, we show how the sensitivity evolves with the exposure. Cyan and orange curves correspond to the sensitivities for normal reactor conditions (i.e., pure $^{235}$U) for $\Delta m^2 = 1$ eV$^2$ and 30 eV$^2$, respectively, whereas the magenta and red curves correspond to lithium loading with $f_{^8\text{Li}} = 0.1$. Dashed curves assume negligible shape systematics, whereas the solid curves introduce a 0.5\% shape uncertainty. Our calculations with negligible shape uncertainty for a 5-ton-year exposure are in rough agreement with the sensitivities in the systematics-dominated limit. Moreover, the uranium-only and lithium-loaded curves asymptote to their ultimate sensitivities in lockstep. From Fig.~\ref{fig:exposureLi}, it is clear that increasing the exposure for pure uranium does not recover the sensitivity to higher $\Delta m^2$ obtained from adding even a small amount of lithium. Simply put, the increase in statistics between 2.0 and 8.0 MeV prompt energy is not enough to compensate for the total absence of events between 8.0 and 12.0 MeV. Given the choice between increasing the exposure of the experiment and increasing the lithium loading of the reactor, the latter is more effective in increasing the ultimate sensitivity.

In summary, our analysis suggests that loading a research core with highly enriched lithium is, from a physics perspective, an attractive possibility for enhancing the sensitivity of a reactor antineutrino experiment to a sterile neutrino, especially for moderately large values of $\Delta m^2$. That said, there are several practical hurdles that make any realization of such a concept technically demanding. 

\section{Conclusions}
\label{sec:con}
\setcounter{equation}{0}

In this paper, we have studied the green-field site optimization of short-baseline reactor experiments for their sensitivity to neutrino oscillations, i.e., to $\overline{\nu}_e$ disappearance. The key to a sensitive experiment is the comparison of neutrino spectra measured at different baselines, which renders the results independent from reactor neutrino flux predictions. We perform the optimization by placing detectors at two baselines simultaneously and by finding the combination of baselines which provides the best overall sensitivity. We also find that the optimized difference of baselines is $\mathcal{O}(1)$ m and thus can, in principle, be accommodated within a single detector with position resolution, which we specifically illustrate with TAO. We look at two types of facilities: commercial reactors, with large, high-power cores and a closest approach of 10-25\,m; and research reactors with compact, low-power cores and a closest approach of 3-6\,m. Experiments at commercial reactors clearly outperform research reactor for $\Delta m^2<1-2\,\mathrm{eV}^2$ under a variety of scenarios, whereas for larger $\Delta m^2>2\,\mathrm{eV}^2$ experiments at research reactors do better. For all cases, getting as close as possible with the best possible energy resolution is valuable. These results are summarized in Fig.~\ref{fig:aggregate}.

We have also investigated how the sensitivity scales with exposure and find that is saturates at around 10 ton-years even for systematics as small as 0.5\%. We further extend the analysis by looking at $^7$Li doping of a reactor to create higher-energy neutrinos. This results in enhanced sensitivity to large values of $\Delta m^2$  ($\sim30\,\mathrm{eV}^2$) even for a modest, and potentially realistic, lithium fraction, of the order 10\%. 

Reactor neutrino experiments can offer a very good sensitivity to oscillations from $\Delta m^2 = 0.1-30\,\mathrm{eV}^2$ with ton-scale experiments, where two or more baselines may be realized in the same detector. Sensitivities below $\sin^22\theta<10^{-2}$ are possible even with reasonable systematics assumptions, pushing towards $\sin^22\theta<10^{-3}$ however seems challenging.  TAO has very good sensitivity but does not compare to a purpose-built experiment because of its relatively large distance to the reactor. Therefore, there is a strong case for a dedicated, short-baseline, commercial-reactor experiment with good energy resolution which currently does not exist. To cover the full range of mass-squared splittings, it is essential to use both research and commercial reactors.

\section*{Acknowledgements}
The work of J.M.B.~is supported by NSF Grant No.~PHY-1630782 and by Heising-Simons Foundation Grant 2017-228. The work of L.D.~and P.H.~is supported by DOE Office of Science Grant No.~\protect{DE-SC00018327}. P.H.~would like to acknowledge useful discussions with A.~Conant. We thank the anonymous referee for bringing the existence of Refs.~\cite{Heeger:2012tc, Heeger:2013ema, Lyashuk:2016lpn} to our attention.


\appendix
\setcounter{equation}{0}

\section{Averaging oscillations over production region}
\label{app:averaging}

If the geometry of an experiment is such that the extent of the core or detector (or both) is not small compared to the distance between them, or if the wavelength associated with a particular oscillation is smaller than either of these, then the expression for the oscillation probability in Eq.~\eqref{eq:prob} must be flux averaged over the production and detection regions. In this case, the proper oscillation probability is formally given by
\begin{equation}
    \label{eq:avg1}
    \langle P_{\overline e \overline e} \rangle = \dfrac{\int d^3 \vec x \, d^3 \vec y \, \, \dfrac{P_{\overline e \overline e}}{|\vec y - \vec x|^2}}{\int d^3 \vec x \, d^3 \vec y \, \,  \dfrac{1}{|\vec y - \vec x|^2}},
\end{equation}
where $\vec x$ and $\vec y$ are the integration variables over the volumes of the core and detector, respectively, and the factor of $\frac{1}{|\vec y - \vec x|^2}$ in either integral accounts for the inverse-square dependence of the flux. For the pseudoexperiments we consider, the detector is assumed pointlike; the integral over $d^3 \vec y$ results in a trivial replacement $\vec y \to \vec L$, the displacement between the center of the core and the detector. Moreover, we approximate the core to be a sphere of radius $R$. Therefore, we may rewrite Eq.~\eqref{eq:avg1} as
\begin{widetext}
\begin{equation}
    \langle P_{\overline e \overline e} \rangle = 1 - \sin^2 2\theta \times \dfrac{\int \dfrac{dr \, d\cos \phi \, \, r^2 \sin^2 \big(q \sqrt{L^2 + r^2 + 2 r L \cos \phi}\big)}{L^2 + r^2 + 2 r L \cos \phi}}{\int \dfrac{dr \, d\cos \phi \, \, r^2}{L^2 + r^2 + 2 r L \cos \phi}} \equiv 1 - \sin^2 2\theta \times F(q),
\end{equation}
where we have employed Eq.~\eqref{eq:prob} and abbreviated $q \equiv \frac{\Delta m^2}{4E_\nu}$. The function $F(q)$ may be evaluated analytically:
\begin{align}
    F(q) & = \dfrac{1}{2} + \Big\{4qR \sin(2qL) \cos(2qR) - 4qL \sin(2qR) \cos(2qL) -2\sin(2qL) \sin(2qR) \Big. \nonumber \\
    & \qquad \Big. + 4q^2(L^2-R^2) \Big[ \text{Ci}\big(2q(L+R)\big) - \text{Ci}\big(2q(L-R)\big)\Big] \Big\} \times \Big\{16q^2RL - 8q^2(L^2-R^2)\ln \left(\frac{L+R}{L-R}\right) \Big\}^{-1},
    \label{eq:geom}
\end{align}
where Ci$(x)$ is the cosine integral function, defined as
\begin{equation}
    \text{Ci}(x) = -\int_x^{\infty} dt \, \frac{\cos(t)}{t}.
\end{equation}
One can verify from Eq.~\eqref{eq:geom} that $F(q)\to \sin^2(q L)$ as $R\to0$.
\end{widetext}


\bibliographystyle{apsrev-title}
\bibliography{references.bib}

\begin{thebibliography}{84}
\expandafter\ifx\csname natexlab\endcsname\relax\def\natexlab#1{#1}\fi
\expandafter\ifx\csname bibnamefont\endcsname\relax
  \def\bibnamefont#1{#1}\fi
\expandafter\ifx\csname bibfnamefont\endcsname\relax
  \def\bibfnamefont#1{#1}\fi
\expandafter\ifx\csname citenamefont\endcsname\relax
  \def\citenamefont#1{#1}\fi
\expandafter\ifx\csname url\endcsname\relax
  \def\url#1{\texttt{#1}}\fi
\expandafter\ifx\csname urlprefix\endcsname\relax\def\urlprefix{URL }\fi
\providecommand{\bibinfo}[2]{#2}
\providecommand{\eprint}[2][]{\url{#2}}

\bibitem[{\citenamefont{Cowan et~al.}(1956)\citenamefont{Cowan, Reines,
  Harrison, Kruse, and McGuire}}]{Cowan:1992xc}
\bibinfo{author}{\bibfnamefont{C.~L.} \bibnamefont{Cowan}},
  \bibinfo{author}{\bibfnamefont{F.}~\bibnamefont{Reines}},
  \bibinfo{author}{\bibfnamefont{F.~B.} \bibnamefont{Harrison}},
  \bibinfo{author}{\bibfnamefont{H.~W.} \bibnamefont{Kruse}}, \bibnamefont{and}
  \bibinfo{author}{\bibfnamefont{A.~D.} \bibnamefont{McGuire}}, ``{Detection of
  the free neutrino: A Confirmation},'' \bibinfo{journal}{Science}
  \textbf{\bibinfo{volume}{124}}, \bibinfo{pages}{103} (\bibinfo{year}{1956}).

\bibitem[{\citenamefont{Eguchi et~al.}(2003)}]{Eguchi:2002dm}
\bibinfo{author}{\bibfnamefont{K.}~\bibnamefont{Eguchi}} \bibnamefont{et~al.}
  (\bibinfo{collaboration}{KamLAND}), ``{First results from KamLAND: Evidence
  for reactor anti-neutrino disappearance},'' \bibinfo{journal}{Phys. Rev.
  Lett.} \textbf{\bibinfo{volume}{90}}, \bibinfo{pages}{021802}
  (\bibinfo{year}{2003}), \eprint{hep-ex/0212021}.

\bibitem[{\citenamefont{Abe et~al.}(2012)}]{Abe:2011fz}
\bibinfo{author}{\bibfnamefont{Y.}~\bibnamefont{Abe}} \bibnamefont{et~al.}
  (\bibinfo{collaboration}{Double Chooz}), ``{Indication of Reactor
  $\bar{\nu}_e$ Disappearance in the Double Chooz Experiment},''
  \bibinfo{journal}{Phys. Rev. Lett.} \textbf{\bibinfo{volume}{108}},
  \bibinfo{pages}{131801} (\bibinfo{year}{2012}), \eprint{1112.6353}.

\bibitem[{\citenamefont{An et~al.}(2012)}]{An:2012eh}
\bibinfo{author}{\bibfnamefont{F.~P.} \bibnamefont{An}} \bibnamefont{et~al.}
  (\bibinfo{collaboration}{Daya Bay}), ``{Observation of electron-antineutrino
  disappearance at Daya Bay},'' \bibinfo{journal}{Phys. Rev. Lett.}
  \textbf{\bibinfo{volume}{108}}, \bibinfo{pages}{171803}
  (\bibinfo{year}{2012}), \eprint{1203.1669}.

\bibitem[{\citenamefont{Ahn et~al.}(2012)}]{Ahn:2012nd}
\bibinfo{author}{\bibfnamefont{J.~K.} \bibnamefont{Ahn}} \bibnamefont{et~al.}
  (\bibinfo{collaboration}{RENO}), ``{Observation of Reactor Electron
  Antineutrino Disappearance in the RENO Experiment},'' \bibinfo{journal}{Phys.
  Rev. Lett.} \textbf{\bibinfo{volume}{108}}, \bibinfo{pages}{191802}
  (\bibinfo{year}{2012}), \eprint{1204.0626}.

\bibitem[{\citenamefont{An et~al.}(2016{\natexlab{a}})}]{An:2015jdp}
\bibinfo{author}{\bibfnamefont{F.}~\bibnamefont{An}} \bibnamefont{et~al.}
  (\bibinfo{collaboration}{JUNO}), ``{Neutrino Physics with JUNO},''
  \bibinfo{journal}{J. Phys. G} \textbf{\bibinfo{volume}{43}},
  \bibinfo{pages}{030401} (\bibinfo{year}{2016}{\natexlab{a}}),
  \eprint{1507.05613}.

\bibitem[{\citenamefont{Mueller et~al.}(2011)}]{Mueller:2011nm}
\bibinfo{author}{\bibfnamefont{{\relax Th}.~A.} \bibnamefont{Mueller}}
  \bibnamefont{et~al.}, ``{Improved Predictions of Reactor Antineutrino
  Spectra},'' \bibinfo{journal}{Phys. Rev.} \textbf{\bibinfo{volume}{C83}},
  \bibinfo{pages}{054615} (\bibinfo{year}{2011}), \eprint{1101.2663}.

\bibitem[{\citenamefont{Huber}(2011)}]{Huber:2011wv}
\bibinfo{author}{\bibfnamefont{P.}~\bibnamefont{Huber}}, ``{On the
  determination of anti-neutrino spectra from nuclear reactors},''
  \bibinfo{journal}{Phys. Rev.} \textbf{\bibinfo{volume}{C84}},
  \bibinfo{pages}{024617} (\bibinfo{year}{2011}), \bibinfo{note}{[Erratum:
  Phys. Rev. C85, 029901 (2012)]}, \eprint{1106.0687}.

\bibitem[{\citenamefont{Mention et~al.}(2011)\citenamefont{Mention, Fechner,
  Lasserre, Mueller, Lhuillier, Cribier, and Letourneau}}]{Mention:2011rk}
\bibinfo{author}{\bibfnamefont{G.}~\bibnamefont{Mention}},
  \bibinfo{author}{\bibfnamefont{M.}~\bibnamefont{Fechner}},
  \bibinfo{author}{\bibfnamefont{{\relax Th}.}~\bibnamefont{Lasserre}},
  \bibinfo{author}{\bibfnamefont{{\relax Th}.~A.} \bibnamefont{Mueller}},
  \bibinfo{author}{\bibfnamefont{D.}~\bibnamefont{Lhuillier}},
  \bibinfo{author}{\bibfnamefont{M.}~\bibnamefont{Cribier}}, \bibnamefont{and}
  \bibinfo{author}{\bibfnamefont{A.}~\bibnamefont{Letourneau}}, ``{The Reactor
  Antineutrino Anomaly},'' \bibinfo{journal}{Phys. Rev.}
  \textbf{\bibinfo{volume}{D83}}, \bibinfo{pages}{073006}
  (\bibinfo{year}{2011}), \eprint{1101.2755}.

\bibitem[{\citenamefont{Abazajian et~al.}(2012)}]{Abazajian:2012ys}
\bibinfo{author}{\bibfnamefont{K.~N.} \bibnamefont{Abazajian}}
  \bibnamefont{et~al.}, ``{Light Sterile Neutrinos: A White Paper},''
  (\bibinfo{year}{2012}), \eprint{1204.5379}.

\bibitem[{\citenamefont{Giunti et~al.}(2017)\citenamefont{Giunti, Ji, Laveder,
  Li, and Littlejohn}}]{Giunti:2017yid}
\bibinfo{author}{\bibfnamefont{C.}~\bibnamefont{Giunti}},
  \bibinfo{author}{\bibfnamefont{X.~P.} \bibnamefont{Ji}},
  \bibinfo{author}{\bibfnamefont{M.}~\bibnamefont{Laveder}},
  \bibinfo{author}{\bibfnamefont{Y.~F.} \bibnamefont{Li}}, \bibnamefont{and}
  \bibinfo{author}{\bibfnamefont{B.~R.} \bibnamefont{Littlejohn}}, ``{Reactor
  Fuel Fraction Information on the Antineutrino Anomaly},''
  \bibinfo{journal}{JHEP} \textbf{\bibinfo{volume}{10}}, \bibinfo{pages}{143}
  (\bibinfo{year}{2017}), \eprint{1708.01133}.

\bibitem[{\citenamefont{Dentler et~al.}(2017)\citenamefont{Dentler,
  Hern\'{a}ndez-Cabezudo, Kopp, Maltoni, and Schwetz}}]{Dentler:2017tkw}
\bibinfo{author}{\bibfnamefont{M.}~\bibnamefont{Dentler}},
  \bibinfo{author}{\bibfnamefont{A.}~\bibnamefont{Hern\'{a}ndez-Cabezudo}},
  \bibinfo{author}{\bibfnamefont{J.}~\bibnamefont{Kopp}},
  \bibinfo{author}{\bibfnamefont{M.}~\bibnamefont{Maltoni}}, \bibnamefont{and}
  \bibinfo{author}{\bibfnamefont{T.}~\bibnamefont{Schwetz}}, ``{Sterile
  neutrinos or flux uncertainties? -- Status of the reactor anti-neutrino
  anomaly},'' \bibinfo{journal}{JHEP} \textbf{\bibinfo{volume}{11}},
  \bibinfo{pages}{099} (\bibinfo{year}{2017}), \eprint{1709.04294}.

\bibitem[{\citenamefont{Dentler et~al.}(2018)\citenamefont{Dentler,
  Hern\'{a}ndez-Cabezudo, Kopp, Machado, Maltoni, Martinez-Soler, and
  Schwetz}}]{Dentler:2018sju}
\bibinfo{author}{\bibfnamefont{M.}~\bibnamefont{Dentler}},
  \bibinfo{author}{\bibfnamefont{A.}~\bibnamefont{Hern\'{a}ndez-Cabezudo}},
  \bibinfo{author}{\bibfnamefont{J.}~\bibnamefont{Kopp}},
  \bibinfo{author}{\bibfnamefont{P.~A.~N.} \bibnamefont{Machado}},
  \bibinfo{author}{\bibfnamefont{M.}~\bibnamefont{Maltoni}},
  \bibinfo{author}{\bibfnamefont{I.}~\bibnamefont{Martinez-Soler}},
  \bibnamefont{and} \bibinfo{author}{\bibfnamefont{T.}~\bibnamefont{Schwetz}},
  ``{Updated Global Analysis of Neutrino Oscillations in the Presence of
  eV-Scale Sterile Neutrinos},'' \bibinfo{journal}{JHEP}
  \textbf{\bibinfo{volume}{08}}, \bibinfo{pages}{010} (\bibinfo{year}{2018}),
  \eprint{1803.10661}.

\bibitem[{\citenamefont{Giunti et~al.}(2019)\citenamefont{Giunti, Li,
  Littlejohn, and Surukuchi}}]{Giunti:2019qlt}
\bibinfo{author}{\bibfnamefont{C.}~\bibnamefont{Giunti}},
  \bibinfo{author}{\bibfnamefont{Y.~F.} \bibnamefont{Li}},
  \bibinfo{author}{\bibfnamefont{B.~R.} \bibnamefont{Littlejohn}},
  \bibnamefont{and} \bibinfo{author}{\bibfnamefont{P.~T.}
  \bibnamefont{Surukuchi}}, ``{Diagnosing the Reactor Antineutrino Anomaly with
  Global Antineutrino Flux Data},'' \bibinfo{journal}{Phys. Rev.}
  \textbf{\bibinfo{volume}{D99}}, \bibinfo{pages}{073005}
  (\bibinfo{year}{2019}), \eprint{1901.01807}.

\bibitem[{\citenamefont{Berryman and Huber}(2021)}]{Berryman:2020agd}
\bibinfo{author}{\bibfnamefont{J.~M.} \bibnamefont{Berryman}} \bibnamefont{and}
  \bibinfo{author}{\bibfnamefont{P.}~\bibnamefont{Huber}}, ``{Sterile Neutrinos
  and the Global Reactor Antineutrino Dataset},'' \bibinfo{journal}{JHEP}
  \textbf{\bibinfo{volume}{01}}, \bibinfo{pages}{167} (\bibinfo{year}{2021}),
  \eprint{2005.01756}.

\bibitem[{\citenamefont{Giunti}(2020)}]{Giunti:2020uhv}
\bibinfo{author}{\bibfnamefont{C.}~\bibnamefont{Giunti}}, ``{Statistical
  Significance of Reactor Antineutrino Active-Sterile Oscillations},''
  \bibinfo{journal}{Phys. Rev. D} \textbf{\bibinfo{volume}{101}},
  \bibinfo{pages}{095025} (\bibinfo{year}{2020}), \eprint{2004.07577}.

\bibitem[{\citenamefont{Giunti et~al.}(2021)\citenamefont{Giunti, Li, Ternes,
  and Xin}}]{Giunti:2021kab}
\bibinfo{author}{\bibfnamefont{C.}~\bibnamefont{Giunti}},
  \bibinfo{author}{\bibfnamefont{Y.~F.} \bibnamefont{Li}},
  \bibinfo{author}{\bibfnamefont{C.~A.} \bibnamefont{Ternes}},
  \bibnamefont{and} \bibinfo{author}{\bibfnamefont{Z.}~\bibnamefont{Xin}},
  ``{Reactor antineutrino anomaly in light of recent flux model refinements},''
   (\bibinfo{year}{2021}), \eprint{2110.06820}.

\bibitem[{\citenamefont{B\"oser et~al.}(2020)\citenamefont{B\"oser, Buck,
  Giunti, Lesgourgues, Ludhova, Mertens, Schukraft, and Wurm}}]{Boser:2019rta}
\bibinfo{author}{\bibfnamefont{S.}~\bibnamefont{B\"oser}},
  \bibinfo{author}{\bibfnamefont{C.}~\bibnamefont{Buck}},
  \bibinfo{author}{\bibfnamefont{C.}~\bibnamefont{Giunti}},
  \bibinfo{author}{\bibfnamefont{J.}~\bibnamefont{Lesgourgues}},
  \bibinfo{author}{\bibfnamefont{L.}~\bibnamefont{Ludhova}},
  \bibinfo{author}{\bibfnamefont{S.}~\bibnamefont{Mertens}},
  \bibinfo{author}{\bibfnamefont{A.}~\bibnamefont{Schukraft}},
  \bibnamefont{and} \bibinfo{author}{\bibfnamefont{M.}~\bibnamefont{Wurm}},
  ``{Status of Light Sterile Neutrino Searches},'' \bibinfo{journal}{Prog.
  Part. Nucl. Phys.} \textbf{\bibinfo{volume}{111}}, \bibinfo{pages}{103736}
  (\bibinfo{year}{2020}), \eprint{1906.01739}.

\bibitem[{\citenamefont{Ko et~al.}(2017)}]{NEOS:2016wee}
\bibinfo{author}{\bibfnamefont{Y.-J.} \bibnamefont{Ko}} \bibnamefont{et~al.}
  (\bibinfo{collaboration}{NEOS}), ``{Sterile Neutrino Search at the NEOS
  Experiment},'' \bibinfo{journal}{Phys. Rev. Lett.}
  \textbf{\bibinfo{volume}{118}}, \bibinfo{pages}{121802}
  (\bibinfo{year}{2017}), \eprint{1610.05134}.

\bibitem[{\citenamefont{Alekseev et~al.}(2018)}]{DANSS:2018fnn}
\bibinfo{author}{\bibfnamefont{I.}~\bibnamefont{Alekseev}} \bibnamefont{et~al.}
  (\bibinfo{collaboration}{DANSS}), ``{Search for sterile neutrinos at the
  DANSS experiment},'' \bibinfo{journal}{Phys. Lett. B}
  \textbf{\bibinfo{volume}{787}}, \bibinfo{pages}{56} (\bibinfo{year}{2018}),
  \eprint{1804.04046}.

\bibitem[{\citenamefont{Serebrov et~al.}(2021)}]{Serebrov:2020kmd}
\bibinfo{author}{\bibfnamefont{A.~P.} \bibnamefont{Serebrov}}
  \bibnamefont{et~al.}, ``{Search for sterile neutrinos with the Neutrino-4
  experiment and measurement results},'' \bibinfo{journal}{Phys. Rev. D}
  \textbf{\bibinfo{volume}{104}}, \bibinfo{pages}{032003}
  (\bibinfo{year}{2021}), \eprint{2005.05301}.

\bibitem[{\citenamefont{Andriamirado
  et~al.}(2021{\natexlab{a}})}]{PROSPECT:2020sxr}
\bibinfo{author}{\bibfnamefont{M.}~\bibnamefont{Andriamirado}}
  \bibnamefont{et~al.} (\bibinfo{collaboration}{PROSPECT}), ``{Improved
  short-baseline neutrino oscillation search and energy spectrum measurement
  with the PROSPECT experiment at HFIR},'' \bibinfo{journal}{Phys. Rev. D}
  \textbf{\bibinfo{volume}{103}}, \bibinfo{pages}{032001}
  (\bibinfo{year}{2021}{\natexlab{a}}), \eprint{2006.11210}.

\bibitem[{\citenamefont{Almaz\'an et~al.}(2021)}]{STEREO:2020hup}
\bibinfo{author}{\bibfnamefont{H.}~\bibnamefont{Almaz\'an}}
  \bibnamefont{et~al.} (\bibinfo{collaboration}{STEREO}), ``{First antineutrino
  energy spectrum from $^{235}$U fissions with the STEREO detector at ILL},''
  \bibinfo{journal}{J. Phys. G} \textbf{\bibinfo{volume}{48}},
  \bibinfo{pages}{075107} (\bibinfo{year}{2021}), \eprint{2010.01876}.

\bibitem[{\citenamefont{Hayes et~al.}(2014)\citenamefont{Hayes, Friar, Garvey,
  Jungman, and Jonkmans}}]{Hayes:2013wra}
\bibinfo{author}{\bibfnamefont{A.~C.} \bibnamefont{Hayes}},
  \bibinfo{author}{\bibfnamefont{J.~L.} \bibnamefont{Friar}},
  \bibinfo{author}{\bibfnamefont{G.~T.} \bibnamefont{Garvey}},
  \bibinfo{author}{\bibfnamefont{G.}~\bibnamefont{Jungman}}, \bibnamefont{and}
  \bibinfo{author}{\bibfnamefont{G.}~\bibnamefont{Jonkmans}}, ``{Systematic
  Uncertainties in the Analysis of the Reactor Neutrino Anomaly},''
  \bibinfo{journal}{Phys. Rev. Lett.} \textbf{\bibinfo{volume}{112}},
  \bibinfo{pages}{202501} (\bibinfo{year}{2014}), \eprint{1309.4146}.

\bibitem[{\citenamefont{Hayen et~al.}(2018)\citenamefont{Hayen, Kostensalo,
  Severijns, and Suhonen}}]{Hayen:2018uyg}
\bibinfo{author}{\bibfnamefont{L.}~\bibnamefont{Hayen}},
  \bibinfo{author}{\bibfnamefont{J.}~\bibnamefont{Kostensalo}},
  \bibinfo{author}{\bibfnamefont{N.}~\bibnamefont{Severijns}},
  \bibnamefont{and} \bibinfo{author}{\bibfnamefont{J.}~\bibnamefont{Suhonen}},
  ``{First forbidden transitions in the reactor anomaly},''
  (\bibinfo{year}{2018}), \eprint{1805.12259}.

\bibitem[{\citenamefont{Estienne et~al.}(2019)}]{Estienne:2019ujo}
\bibinfo{author}{\bibfnamefont{M.}~\bibnamefont{Estienne}}
  \bibnamefont{et~al.}, ``{Updated Summation Model: An Improved Agreement with
  the Daya Bay Antineutrino Fluxes},'' \bibinfo{journal}{Phys. Rev. Lett.}
  \textbf{\bibinfo{volume}{123}}, \bibinfo{pages}{022502}
  (\bibinfo{year}{2019}), \eprint{1904.09358}.

\bibitem[{\citenamefont{Hayen et~al.}(2019)\citenamefont{Hayen, Kostensalo,
  Severijns, and Suhonen}}]{Hayen:2019eop}
\bibinfo{author}{\bibfnamefont{L.}~\bibnamefont{Hayen}},
  \bibinfo{author}{\bibfnamefont{J.}~\bibnamefont{Kostensalo}},
  \bibinfo{author}{\bibfnamefont{N.}~\bibnamefont{Severijns}},
  \bibnamefont{and} \bibinfo{author}{\bibfnamefont{J.}~\bibnamefont{Suhonen}},
  ``{First-forbidden transitions in the reactor anomaly},''
  \bibinfo{journal}{Phys. Rev. C} \textbf{\bibinfo{volume}{100}},
  \bibinfo{pages}{054323} (\bibinfo{year}{2019}), \eprint{1908.08302}.

\bibitem[{\citenamefont{An et~al.}(2017)}]{DayaBay:2017jkb}
\bibinfo{author}{\bibfnamefont{F.~P.} \bibnamefont{An}} \bibnamefont{et~al.}
  (\bibinfo{collaboration}{Daya Bay}), ``{Evolution of the Reactor Antineutrino
  Flux and Spectrum at Daya Bay},'' \bibinfo{journal}{Phys. Rev. Lett.}
  \textbf{\bibinfo{volume}{118}}, \bibinfo{pages}{251801}
  (\bibinfo{year}{2017}), \eprint{1704.01082}.

\bibitem[{\citenamefont{Giunti}(2017)}]{Giunti:2017nww}
\bibinfo{author}{\bibfnamefont{C.}~\bibnamefont{Giunti}}, ``{Improved
  Determination of the $^{235}\text{U}$ and $^{239}\text{Pu}$ Reactor
  Antineutrino Cross Sections per Fission},'' \bibinfo{journal}{Phys. Rev. D}
  \textbf{\bibinfo{volume}{96}}, \bibinfo{pages}{033005}
  (\bibinfo{year}{2017}), \eprint{1704.02276}.

\bibitem[{\citenamefont{Gebre et~al.}(2018)\citenamefont{Gebre, Littlejohn, and
  Surukuchi}}]{Gebre:2017vmm}
\bibinfo{author}{\bibfnamefont{Y.}~\bibnamefont{Gebre}},
  \bibinfo{author}{\bibfnamefont{B.}~\bibnamefont{Littlejohn}},
  \bibnamefont{and}
  \bibinfo{author}{\bibfnamefont{P.}~\bibnamefont{Surukuchi}}, ``{Prospects for
  Improved Understanding of Isotopic Reactor Antineutrino Fluxes},''
  \bibinfo{journal}{Phys. Rev. D} \textbf{\bibinfo{volume}{97}},
  \bibinfo{pages}{013003} (\bibinfo{year}{2018}), \eprint{1709.10051}.

\bibitem[{\citenamefont{Bak et~al.}(2019)}]{RENO:2018pwo}
\bibinfo{author}{\bibfnamefont{G.}~\bibnamefont{Bak}} \bibnamefont{et~al.}
  (\bibinfo{collaboration}{RENO}), ``{Fuel-composition dependent reactor
  antineutrino yield at RENO},'' \bibinfo{journal}{Phys. Rev. Lett.}
  \textbf{\bibinfo{volume}{122}}, \bibinfo{pages}{232501}
  (\bibinfo{year}{2019}), \eprint{1806.00574}.

\bibitem[{\citenamefont{Berryman and Huber}(2020)}]{Berryman:2019hme}
\bibinfo{author}{\bibfnamefont{J.~M.} \bibnamefont{Berryman}} \bibnamefont{and}
  \bibinfo{author}{\bibfnamefont{P.}~\bibnamefont{Huber}}, ``{Reevaluating
  Reactor Antineutrino Anomalies with Updated Flux Predictions},''
  \bibinfo{journal}{Phys. Rev. D} \textbf{\bibinfo{volume}{101}},
  \bibinfo{pages}{015008} (\bibinfo{year}{2020}), \eprint{1909.09267}.

\bibitem[{\citenamefont{Kopeikin et~al.}(2021)\citenamefont{Kopeikin,
  Skorokhvatov, and Titov}}]{Kopeikin:2021ugh}
\bibinfo{author}{\bibfnamefont{V.}~\bibnamefont{Kopeikin}},
  \bibinfo{author}{\bibfnamefont{M.}~\bibnamefont{Skorokhvatov}},
  \bibnamefont{and} \bibinfo{author}{\bibfnamefont{O.}~\bibnamefont{Titov}},
  ``{Reevaluating reactor antineutrino spectra with new measurements of the
  ratio between U235 and Pu239 \ensuremath{\beta} spectra},''
  \bibinfo{journal}{Phys. Rev. D} \textbf{\bibinfo{volume}{104}},
  \bibinfo{pages}{L071301} (\bibinfo{year}{2021}), \eprint{2103.01684}.

\bibitem[{\citenamefont{Abe et~al.}(2014)}]{DoubleChooz:2014kuw}
\bibinfo{author}{\bibfnamefont{Y.}~\bibnamefont{Abe}} \bibnamefont{et~al.}
  (\bibinfo{collaboration}{Double Chooz}), ``{Improved measurements of the
  neutrino mixing angle $\theta_{13}$ with the Double Chooz detector},''
  \bibinfo{journal}{JHEP} \textbf{\bibinfo{volume}{10}}, \bibinfo{pages}{086}
  (\bibinfo{year}{2014}), \bibinfo{note}{[Erratum: JHEP 02, 074 (2015)]},
  \eprint{1406.7763}.

\bibitem[{\citenamefont{An et~al.}(2016{\natexlab{b}})}]{DayaBay:2015lja}
\bibinfo{author}{\bibfnamefont{F.~P.} \bibnamefont{An}} \bibnamefont{et~al.}
  (\bibinfo{collaboration}{Daya Bay}), ``{Measurement of the Reactor
  Antineutrino Flux and Spectrum at Daya Bay},'' \bibinfo{journal}{Phys. Rev.
  Lett.} \textbf{\bibinfo{volume}{116}}, \bibinfo{pages}{061801}
  (\bibinfo{year}{2016}{\natexlab{b}}), \bibinfo{note}{[Erratum: Phys.Rev.Lett.
  118, 099902 (2017)]}, \eprint{1508.04233}.

\bibitem[{\citenamefont{Choi et~al.}(2016)}]{RENO:2015ksa}
\bibinfo{author}{\bibfnamefont{J.~H.} \bibnamefont{Choi}} \bibnamefont{et~al.}
  (\bibinfo{collaboration}{RENO}), ``{Observation of Energy and Baseline
  Dependent Reactor Antineutrino Disappearance in the RENO Experiment},''
  \bibinfo{journal}{Phys. Rev. Lett.} \textbf{\bibinfo{volume}{116}},
  \bibinfo{pages}{211801} (\bibinfo{year}{2016}), \eprint{1511.05849}.

\bibitem[{\citenamefont{Hayes et~al.}(2015)\citenamefont{Hayes, Friar, Garvey,
  Ibeling, Jungman, Kawano, and Mills}}]{Hayes:2015yka}
\bibinfo{author}{\bibfnamefont{A.~C.} \bibnamefont{Hayes}},
  \bibinfo{author}{\bibfnamefont{J.~L.} \bibnamefont{Friar}},
  \bibinfo{author}{\bibfnamefont{G.~T.} \bibnamefont{Garvey}},
  \bibinfo{author}{\bibfnamefont{D.}~\bibnamefont{Ibeling}},
  \bibinfo{author}{\bibfnamefont{G.}~\bibnamefont{Jungman}},
  \bibinfo{author}{\bibfnamefont{T.}~\bibnamefont{Kawano}}, \bibnamefont{and}
  \bibinfo{author}{\bibfnamefont{R.~W.} \bibnamefont{Mills}}, ``{Possible
  origins and implications of the shoulder in reactor neutrino spectra},''
  \bibinfo{journal}{Phys. Rev. D} \textbf{\bibinfo{volume}{92}},
  \bibinfo{pages}{033015} (\bibinfo{year}{2015}), \eprint{1506.00583}.

\bibitem[{\citenamefont{Sonzogni et~al.}(2017)\citenamefont{Sonzogni,
  McCutchan, and Hayes}}]{Sonzogni:2017wxy}
\bibinfo{author}{\bibfnamefont{A.~A.} \bibnamefont{Sonzogni}},
  \bibinfo{author}{\bibfnamefont{E.~A.} \bibnamefont{McCutchan}},
  \bibnamefont{and} \bibinfo{author}{\bibfnamefont{A.~C.} \bibnamefont{Hayes}},
  ``{Dissecting Reactor Antineutrino Flux Calculations},''
  \bibinfo{journal}{Phys. Rev. Lett.} \textbf{\bibinfo{volume}{119}},
  \bibinfo{pages}{112501} (\bibinfo{year}{2017}).

\bibitem[{\citenamefont{Mention et~al.}(2017)\citenamefont{Mention, Vivier,
  Gaffiot, Lasserre, Letourneau, and Materna}}]{Mention:2017dyq}
\bibinfo{author}{\bibfnamefont{G.}~\bibnamefont{Mention}},
  \bibinfo{author}{\bibfnamefont{M.}~\bibnamefont{Vivier}},
  \bibinfo{author}{\bibfnamefont{J.}~\bibnamefont{Gaffiot}},
  \bibinfo{author}{\bibfnamefont{T.}~\bibnamefont{Lasserre}},
  \bibinfo{author}{\bibfnamefont{A.}~\bibnamefont{Letourneau}},
  \bibnamefont{and} \bibinfo{author}{\bibfnamefont{T.}~\bibnamefont{Materna}},
  ``{Reactor antineutrino shoulder explained by energy scale
  nonlinearities?},'' \bibinfo{journal}{Phys. Lett. B}
  \textbf{\bibinfo{volume}{773}}, \bibinfo{pages}{307} (\bibinfo{year}{2017}),
  \eprint{1705.09434}.

\bibitem[{\citenamefont{Littlejohn et~al.}(2018)\citenamefont{Littlejohn,
  Conant, Dwyer, Erickson, Gustafson, and Hermanek}}]{Littlejohn:2018hqm}
\bibinfo{author}{\bibfnamefont{B.~R.} \bibnamefont{Littlejohn}},
  \bibinfo{author}{\bibfnamefont{A.}~\bibnamefont{Conant}},
  \bibinfo{author}{\bibfnamefont{D.~A.} \bibnamefont{Dwyer}},
  \bibinfo{author}{\bibfnamefont{A.}~\bibnamefont{Erickson}},
  \bibinfo{author}{\bibfnamefont{I.}~\bibnamefont{Gustafson}},
  \bibnamefont{and} \bibinfo{author}{\bibfnamefont{K.}~\bibnamefont{Hermanek}},
  ``{Impact of Fission Neutron Energies on Reactor Antineutrino Spectra},''
  \bibinfo{journal}{Phys. Rev. D} \textbf{\bibinfo{volume}{97}},
  \bibinfo{pages}{073007} (\bibinfo{year}{2018}), \eprint{1803.01787}.

\bibitem[{\citenamefont{Berryman et~al.}(2019)\citenamefont{Berryman, Brdar,
  and Huber}}]{Berryman:2018jxt}
\bibinfo{author}{\bibfnamefont{J.~M.} \bibnamefont{Berryman}},
  \bibinfo{author}{\bibfnamefont{V.}~\bibnamefont{Brdar}}, \bibnamefont{and}
  \bibinfo{author}{\bibfnamefont{P.}~\bibnamefont{Huber}}, ``{Particle physics
  origin of the 5 MeV bump in the reactor antineutrino spectrum?},''
  \bibinfo{journal}{Phys. Rev.} \textbf{\bibinfo{volume}{D99}},
  \bibinfo{pages}{055045} (\bibinfo{year}{2019}), \eprint{1803.08506}.

\bibitem[{\citenamefont{Moulai et~al.}(2020)\citenamefont{Moulai, Arg\"uelles,
  Collin, Conrad, Diaz, and Shaevitz}}]{Moulai:2019gpi}
\bibinfo{author}{\bibfnamefont{M.~H.} \bibnamefont{Moulai}},
  \bibinfo{author}{\bibfnamefont{C.~A.} \bibnamefont{Arg\"uelles}},
  \bibinfo{author}{\bibfnamefont{G.~H.} \bibnamefont{Collin}},
  \bibinfo{author}{\bibfnamefont{J.~M.} \bibnamefont{Conrad}},
  \bibinfo{author}{\bibfnamefont{A.}~\bibnamefont{Diaz}}, \bibnamefont{and}
  \bibinfo{author}{\bibfnamefont{M.~H.} \bibnamefont{Shaevitz}}, ``{Combining
  Sterile Neutrino Fits to Short Baseline Data with IceCube Data},''
  \bibinfo{journal}{Phys. Rev. D} \textbf{\bibinfo{volume}{101}},
  \bibinfo{pages}{055020} (\bibinfo{year}{2020}), \eprint{1910.13456}.

\bibitem[{\citenamefont{Adamson et~al.}(2020)}]{Adamson:2020jvo}
\bibinfo{author}{\bibfnamefont{P.}~\bibnamefont{Adamson}} \bibnamefont{et~al.}
  (\bibinfo{collaboration}{MINOS+, Daya Bay}), ``{Improved Constraints on
  Sterile Neutrino Mixing from Disappearance Searches in the MINOS, MINOS+,
  Daya Bay, and Bugey-3 Experiments},'' \bibinfo{journal}{Phys. Rev. Lett.}
  \textbf{\bibinfo{volume}{125}}, \bibinfo{pages}{071801}
  (\bibinfo{year}{2020}), \eprint{2002.00301}.

\bibitem[{\citenamefont{Berryman}(2019)}]{Berryman:2019nvr}
\bibinfo{author}{\bibfnamefont{J.~M.} \bibnamefont{Berryman}}, ``{Constraining
  Sterile Neutrino Cosmology with Terrestrial Oscillation Experiments},''
  \bibinfo{journal}{Phys. Rev. D} \textbf{\bibinfo{volume}{100}},
  \bibinfo{pages}{023540} (\bibinfo{year}{2019}), \eprint{1905.03254}.

\bibitem[{\citenamefont{Gariazzo et~al.}(2019)\citenamefont{Gariazzo, de~Salas,
  and Pastor}}]{Gariazzo:2019gyi}
\bibinfo{author}{\bibfnamefont{S.}~\bibnamefont{Gariazzo}},
  \bibinfo{author}{\bibfnamefont{P.~F.} \bibnamefont{de~Salas}},
  \bibnamefont{and} \bibinfo{author}{\bibfnamefont{S.}~\bibnamefont{Pastor}},
  ``{Thermalisation of sterile neutrinos in the early Universe in the 3+1
  scheme with full mixing matrix},'' \bibinfo{journal}{JCAP}
  \textbf{\bibinfo{volume}{07}}, \bibinfo{pages}{014} (\bibinfo{year}{2019}),
  \eprint{1905.11290}.

\bibitem[{\citenamefont{Hagstotz et~al.}(2021)\citenamefont{Hagstotz, de~Salas,
  Gariazzo, Gerbino, Lattanzi, Vagnozzi, Freese, and
  Pastor}}]{Hagstotz:2020ukm}
\bibinfo{author}{\bibfnamefont{S.}~\bibnamefont{Hagstotz}},
  \bibinfo{author}{\bibfnamefont{P.~F.} \bibnamefont{de~Salas}},
  \bibinfo{author}{\bibfnamefont{S.}~\bibnamefont{Gariazzo}},
  \bibinfo{author}{\bibfnamefont{M.}~\bibnamefont{Gerbino}},
  \bibinfo{author}{\bibfnamefont{M.}~\bibnamefont{Lattanzi}},
  \bibinfo{author}{\bibfnamefont{S.}~\bibnamefont{Vagnozzi}},
  \bibinfo{author}{\bibfnamefont{K.}~\bibnamefont{Freese}}, \bibnamefont{and}
  \bibinfo{author}{\bibfnamefont{S.}~\bibnamefont{Pastor}}, ``{Bounds on light
  sterile neutrino mass and mixing from cosmology and laboratory searches},''
  \bibinfo{journal}{Phys. Rev. D} \textbf{\bibinfo{volume}{104}},
  \bibinfo{pages}{123524} (\bibinfo{year}{2021}), \eprint{2003.02289}.

\bibitem[{\citenamefont{Adams et~al.}(2020)\citenamefont{Adams, Bezrukov,
  Elvin-Poole, Evans, Guzowski, Fearraigh, and
  S\"oldner-Rembold}}]{Adams:2020nue}
\bibinfo{author}{\bibfnamefont{M.}~\bibnamefont{Adams}},
  \bibinfo{author}{\bibfnamefont{F.}~\bibnamefont{Bezrukov}},
  \bibinfo{author}{\bibfnamefont{J.}~\bibnamefont{Elvin-Poole}},
  \bibinfo{author}{\bibfnamefont{J.~J.} \bibnamefont{Evans}},
  \bibinfo{author}{\bibfnamefont{P.}~\bibnamefont{Guzowski}},
  \bibinfo{author}{\bibfnamefont{B.~O.} \bibnamefont{Fearraigh}},
  \bibnamefont{and}
  \bibinfo{author}{\bibfnamefont{S.}~\bibnamefont{S\"oldner-Rembold}},
  ``{Direct comparison of sterile neutrino constraints from cosmological data,
  $\nu_{e}$ disappearance data and $\nu_{\mu}\rightarrow\nu_{e}$ appearance
  data in a $3+1$ model},'' \bibinfo{journal}{Eur. Phys. J. C}
  \textbf{\bibinfo{volume}{80}}, \bibinfo{pages}{758} (\bibinfo{year}{2020}),
  \eprint{2002.07762}.

\bibitem[{\citenamefont{Anselmann et~al.}(1995)}]{GALLEX:1994rym}
\bibinfo{author}{\bibfnamefont{P.}~\bibnamefont{Anselmann}}
  \bibnamefont{et~al.} (\bibinfo{collaboration}{GALLEX}), ``{First results from
  the Cr-51 neutrino source experiment with the GALLEX detector},''
  \bibinfo{journal}{Phys. Lett. B} \textbf{\bibinfo{volume}{342}},
  \bibinfo{pages}{440} (\bibinfo{year}{1995}).

\bibitem[{\citenamefont{Abdurashitov et~al.}(1996)}]{Abdurashitov:1996dp}
\bibinfo{author}{\bibfnamefont{D.~N.} \bibnamefont{Abdurashitov}}
  \bibnamefont{et~al.}, ``{The Russian-American gallium experiment (SAGE) Cr
  neutrino source measurement},'' \bibinfo{journal}{Phys. Rev. Lett.}
  \textbf{\bibinfo{volume}{77}}, \bibinfo{pages}{4708} (\bibinfo{year}{1996}).

\bibitem[{\citenamefont{Hampel et~al.}(1998)}]{GALLEX:1997lja}
\bibinfo{author}{\bibfnamefont{W.}~\bibnamefont{Hampel}} \bibnamefont{et~al.}
  (\bibinfo{collaboration}{GALLEX}), ``{Final results of the Cr-51 neutrino
  source experiments in GALLEX},'' \bibinfo{journal}{Phys. Lett. B}
  \textbf{\bibinfo{volume}{420}}, \bibinfo{pages}{114} (\bibinfo{year}{1998}).

\bibitem[{\citenamefont{Abdurashitov et~al.}(1999)}]{SAGE:1998fvr}
\bibinfo{author}{\bibfnamefont{J.~N.} \bibnamefont{Abdurashitov}}
  \bibnamefont{et~al.} (\bibinfo{collaboration}{SAGE}), ``{Measurement of the
  response of the Russian-American gallium experiment to neutrinos from a Cr-51
  source},'' \bibinfo{journal}{Phys. Rev. C} \textbf{\bibinfo{volume}{59}},
  \bibinfo{pages}{2246} (\bibinfo{year}{1999}), \eprint{hep-ph/9803418}.

\bibitem[{\citenamefont{Abdurashitov et~al.}(2006)}]{Abdurashitov:2005tb}
\bibinfo{author}{\bibfnamefont{J.~N.} \bibnamefont{Abdurashitov}}
  \bibnamefont{et~al.}, ``{Measurement of the response of a Ga solar neutrino
  experiment to neutrinos from an Ar-37 source},'' \bibinfo{journal}{Phys. Rev.
  C} \textbf{\bibinfo{volume}{73}}, \bibinfo{pages}{045805}
  (\bibinfo{year}{2006}), \eprint{nucl-ex/0512041}.

\bibitem[{\citenamefont{Grieb et~al.}(2007)\citenamefont{Grieb, Link, and
  Raghavan}}]{Grieb:2006mp}
\bibinfo{author}{\bibfnamefont{C.}~\bibnamefont{Grieb}},
  \bibinfo{author}{\bibfnamefont{J.}~\bibnamefont{Link}}, \bibnamefont{and}
  \bibinfo{author}{\bibfnamefont{R.~S.} \bibnamefont{Raghavan}}, ``{Probing
  active to sterile neutrino oscillations in the LENS detector},''
  \bibinfo{journal}{Phys. Rev. D} \textbf{\bibinfo{volume}{75}},
  \bibinfo{pages}{093006} (\bibinfo{year}{2007}), \eprint{hep-ph/0611178}.

\bibitem[{\citenamefont{Kaether et~al.}(2010)\citenamefont{Kaether, Hampel,
  Heusser, Kiko, and Kirsten}}]{Kaether:2010ag}
\bibinfo{author}{\bibfnamefont{F.}~\bibnamefont{Kaether}},
  \bibinfo{author}{\bibfnamefont{W.}~\bibnamefont{Hampel}},
  \bibinfo{author}{\bibfnamefont{G.}~\bibnamefont{Heusser}},
  \bibinfo{author}{\bibfnamefont{J.}~\bibnamefont{Kiko}}, \bibnamefont{and}
  \bibinfo{author}{\bibfnamefont{T.}~\bibnamefont{Kirsten}}, ``{Reanalysis of
  the GALLEX solar neutrino flux and source experiments},''
  \bibinfo{journal}{Phys. Lett. B} \textbf{\bibinfo{volume}{685}},
  \bibinfo{pages}{47} (\bibinfo{year}{2010}), \eprint{1001.2731}.

\bibitem[{\citenamefont{Bellini et~al.}(2013)}]{Borexino:2013xxa}
\bibinfo{author}{\bibfnamefont{G.}~\bibnamefont{Bellini}} \bibnamefont{et~al.}
  (\bibinfo{collaboration}{Borexino}), ``{SOX: Short distance neutrino
  Oscillations with BoreXino},'' \bibinfo{journal}{JHEP}
  \textbf{\bibinfo{volume}{08}}, \bibinfo{pages}{038} (\bibinfo{year}{2013}),
  \eprint{1304.7721}.

\bibitem[{\citenamefont{Barinov et~al.}(2021)}]{Barinov:2021asz}
\bibinfo{author}{\bibfnamefont{V.~V.} \bibnamefont{Barinov}}
  \bibnamefont{et~al.}, ``{Results from the Baksan Experiment on Sterile
  Transitions (BEST)},''  (\bibinfo{year}{2021}), \eprint{2109.11482}.

\bibitem[{\citenamefont{Barinov and Gorbunov}(2021)}]{Barinov:2021mjj}
\bibinfo{author}{\bibfnamefont{V.}~\bibnamefont{Barinov}} \bibnamefont{and}
  \bibinfo{author}{\bibfnamefont{D.}~\bibnamefont{Gorbunov}}, ``{BEST Impact on
  Sterile Neutrino Hypothesis},''  (\bibinfo{year}{2021}), \eprint{2109.14654}.

\bibitem[{\citenamefont{Goldhagen et~al.}(2021)\citenamefont{Goldhagen,
  Maltoni, Reichard, and Schwetz}}]{Goldhagen:2021kxe}
\bibinfo{author}{\bibfnamefont{K.}~\bibnamefont{Goldhagen}},
  \bibinfo{author}{\bibfnamefont{M.}~\bibnamefont{Maltoni}},
  \bibinfo{author}{\bibfnamefont{S.}~\bibnamefont{Reichard}}, \bibnamefont{and}
  \bibinfo{author}{\bibfnamefont{T.}~\bibnamefont{Schwetz}}, ``{Testing sterile
  neutrino mixing with present and future solar neutrino data},''
  (\bibinfo{year}{2021}), \eprint{2109.14898}.

\bibitem[{\citenamefont{Albright et~al.}(2004)}]{NeutrinoFactory:2004odt}
\bibinfo{author}{\bibfnamefont{C.}~\bibnamefont{Albright}} \bibnamefont{et~al.}
  (\bibinfo{collaboration}{Neutrino Factory, Muon Collider}), ``{The neutrino
  factory and beta beam experiments and development},''
  (\bibinfo{year}{2004}), \eprint{physics/0411123}.

\bibitem[{\citenamefont{Agarwalla et~al.}(2010)\citenamefont{Agarwalla, Huber,
  and Link}}]{Agarwalla:2009em}
\bibinfo{author}{\bibfnamefont{S.~K.} \bibnamefont{Agarwalla}},
  \bibinfo{author}{\bibfnamefont{P.}~\bibnamefont{Huber}}, \bibnamefont{and}
  \bibinfo{author}{\bibfnamefont{J.~M.} \bibnamefont{Link}}, ``{Constraining
  sterile neutrinos with a low energy beta-beam},'' \bibinfo{journal}{JHEP}
  \textbf{\bibinfo{volume}{01}}, \bibinfo{pages}{071} (\bibinfo{year}{2010}),
  \eprint{0907.3145}.

\bibitem[{\citenamefont{Delgadillo and Huber}(2021)}]{Delgadillo:2020uvm}
\bibinfo{author}{\bibfnamefont{L.}~\bibnamefont{Delgadillo}} \bibnamefont{and}
  \bibinfo{author}{\bibfnamefont{P.}~\bibnamefont{Huber}}, ``{Sterile neutrino
  searches at tagged kaon beams},'' \bibinfo{journal}{Phys. Rev. D}
  \textbf{\bibinfo{volume}{103}}, \bibinfo{pages}{035018}
  (\bibinfo{year}{2021}), \eprint{2010.10268}.

\bibitem[{\citenamefont{Conrad et~al.}(2014)\citenamefont{Conrad, Shaevitz,
  Shimizu, Spitz, Toups, and Winslow}}]{Conrad:2013sqa}
\bibinfo{author}{\bibfnamefont{J.~M.} \bibnamefont{Conrad}},
  \bibinfo{author}{\bibfnamefont{M.~H.} \bibnamefont{Shaevitz}},
  \bibinfo{author}{\bibfnamefont{I.}~\bibnamefont{Shimizu}},
  \bibinfo{author}{\bibfnamefont{J.}~\bibnamefont{Spitz}},
  \bibinfo{author}{\bibfnamefont{M.}~\bibnamefont{Toups}}, \bibnamefont{and}
  \bibinfo{author}{\bibfnamefont{L.}~\bibnamefont{Winslow}}, ``{Precision
  $\bar{\nu}_e$-electron scattering measurements with IsoDAR to search for new
  physics},'' \bibinfo{journal}{Phys. Rev. D} \textbf{\bibinfo{volume}{89}},
  \bibinfo{pages}{072010} (\bibinfo{year}{2014}), \eprint{1307.5081}.

\bibitem[{\citenamefont{Adey et~al.}(2014)}]{Adey:2014rfv}
\bibinfo{author}{\bibfnamefont{D.}~\bibnamefont{Adey}} \bibnamefont{et~al.}
  (\bibinfo{collaboration}{nuSTORM}), ``{Light sterile neutrino sensitivity at
  the nuSTORM facility},'' \bibinfo{journal}{Phys. Rev. D}
  \textbf{\bibinfo{volume}{89}}, \bibinfo{pages}{071301}
  (\bibinfo{year}{2014}), \eprint{1402.5250}.

\bibitem[{\citenamefont{Heeger et~al.}(2013{\natexlab{a}})\citenamefont{Heeger,
  Littlejohn, Mumm, and Tobin}}]{Heeger:2012tc}
\bibinfo{author}{\bibfnamefont{K.~M.} \bibnamefont{Heeger}},
  \bibinfo{author}{\bibfnamefont{B.~R.} \bibnamefont{Littlejohn}},
  \bibinfo{author}{\bibfnamefont{H.~P.} \bibnamefont{Mumm}}, \bibnamefont{and}
  \bibinfo{author}{\bibfnamefont{M.~N.} \bibnamefont{Tobin}}, ``{Experimental
  Parameters for a Reactor Antineutrino Experiment at Very Short Baselines},''
  \bibinfo{journal}{Phys. Rev. D} \textbf{\bibinfo{volume}{87}},
  \bibinfo{pages}{073008} (\bibinfo{year}{2013}{\natexlab{a}}),
  \eprint{1212.2182}.

\bibitem[{\citenamefont{Heeger et~al.}(2013{\natexlab{b}})\citenamefont{Heeger,
  Littlejohn, and Mumm}}]{Heeger:2013ema}
\bibinfo{author}{\bibfnamefont{K.~M.} \bibnamefont{Heeger}},
  \bibinfo{author}{\bibfnamefont{B.~R.} \bibnamefont{Littlejohn}},
  \bibnamefont{and} \bibinfo{author}{\bibfnamefont{H.~P.} \bibnamefont{Mumm}},
  ``{Multiple Detectors for a Short-Baseline Neutrino Oscillation Search Near
  Reactors},''  (\bibinfo{year}{2013}{\natexlab{b}}), \eprint{1307.2859}.

\bibitem[{\citenamefont{Bak et~al.}(2018)}]{RENO:2018dro}
\bibinfo{author}{\bibfnamefont{G.}~\bibnamefont{Bak}} \bibnamefont{et~al.}
  (\bibinfo{collaboration}{RENO}), ``{Measurement of Reactor Antineutrino
  Oscillation Amplitude and Frequency at RENO},'' \bibinfo{journal}{Phys. Rev.
  Lett.} \textbf{\bibinfo{volume}{121}}, \bibinfo{pages}{201801}
  (\bibinfo{year}{2018}), \eprint{1806.00248}.

\bibitem[{\citenamefont{Adey et~al.}(2018)}]{DayaBay:2018yms}
\bibinfo{author}{\bibfnamefont{D.}~\bibnamefont{Adey}} \bibnamefont{et~al.}
  (\bibinfo{collaboration}{Daya Bay}), ``{Measurement of the Electron
  Antineutrino Oscillation with 1958 Days of Operation at Daya Bay},''
  \bibinfo{journal}{Phys. Rev. Lett.} \textbf{\bibinfo{volume}{121}},
  \bibinfo{pages}{241805} (\bibinfo{year}{2018}), \eprint{1809.02261}.

\bibitem[{\citenamefont{de~Kerret et~al.}(2020)}]{DoubleChooz:2019qbj}
\bibinfo{author}{\bibfnamefont{H.}~\bibnamefont{de~Kerret}}
  \bibnamefont{et~al.} (\bibinfo{collaboration}{Double Chooz}), ``{Double Chooz
  $\theta_{13}$ measurement via total neutron capture detection},''
  \bibinfo{journal}{Nature Phys.} \textbf{\bibinfo{volume}{16}},
  \bibinfo{pages}{558} (\bibinfo{year}{2020}), \eprint{1901.09445}.

\bibitem[{\citenamefont{Andriamirado
  et~al.}(2021{\natexlab{b}})}]{Andriamirado:2020erz}
\bibinfo{author}{\bibfnamefont{M.}~\bibnamefont{Andriamirado}}
  \bibnamefont{et~al.} (\bibinfo{collaboration}{PROSPECT}), ``{Improved
  short-baseline neutrino oscillation search and energy spectrum measurement
  with the PROSPECT experiment at HFIR},'' \bibinfo{journal}{Phys. Rev. D}
  \textbf{\bibinfo{volume}{103}}, \bibinfo{pages}{032001}
  (\bibinfo{year}{2021}{\natexlab{b}}), \eprint{2006.11210}.

\bibitem[{\citenamefont{Abusleme et~al.}(2020)}]{Abusleme:2020bzt}
\bibinfo{author}{\bibfnamefont{A.}~\bibnamefont{Abusleme}} \bibnamefont{et~al.}
  (\bibinfo{collaboration}{JUNO}), ``{TAO Conceptual Design Report: A Precision
  Measurement of the Reactor Antineutrino Spectrum with Sub-percent Energy
  Resolution},''  (\bibinfo{year}{2020}), \eprint{2005.08745}.

\bibitem[{\citenamefont{Almaz\'an et~al.}(2020)}]{STEREO:2019ztb}
\bibinfo{author}{\bibfnamefont{H.}~\bibnamefont{Almaz\'an}}
  \bibnamefont{et~al.} (\bibinfo{collaboration}{STEREO}), ``{Improved sterile
  neutrino constraints from the STEREO experiment with 179 days of reactor-on
  data},'' \bibinfo{journal}{Phys. Rev. D} \textbf{\bibinfo{volume}{102}},
  \bibinfo{pages}{052002} (\bibinfo{year}{2020}), \eprint{1912.06582}.

\bibitem[{\citenamefont{Agostini and Neumair}(2020)}]{Agostini:2019jup}
\bibinfo{author}{\bibfnamefont{M.}~\bibnamefont{Agostini}} \bibnamefont{and}
  \bibinfo{author}{\bibfnamefont{B.}~\bibnamefont{Neumair}}, ``{Statistical
  Methods Applied to the Search of Sterile Neutrinos},'' \bibinfo{journal}{Eur.
  Phys. J. C} \textbf{\bibinfo{volume}{80}}, \bibinfo{pages}{750}
  (\bibinfo{year}{2020}), \eprint{1906.11854}.

\bibitem[{\citenamefont{Coloma et~al.}(2021)\citenamefont{Coloma, Huber, and
  Schwetz}}]{Coloma:2020ajw}
\bibinfo{author}{\bibfnamefont{P.}~\bibnamefont{Coloma}},
  \bibinfo{author}{\bibfnamefont{P.}~\bibnamefont{Huber}}, \bibnamefont{and}
  \bibinfo{author}{\bibfnamefont{T.}~\bibnamefont{Schwetz}}, ``{Statistical
  interpretation of sterile neutrino oscillation searches at reactors},''
  \bibinfo{journal}{Eur. Phys. J. C} \textbf{\bibinfo{volume}{81}},
  \bibinfo{pages}{2} (\bibinfo{year}{2021}), \eprint{2008.06083}.

\bibitem[{\citenamefont{Qian et~al.}(2016)\citenamefont{Qian, Tan, Ling,
  Nakajima, and Zhang}}]{Qian:2014nha}
\bibinfo{author}{\bibfnamefont{X.}~\bibnamefont{Qian}},
  \bibinfo{author}{\bibfnamefont{A.}~\bibnamefont{Tan}},
  \bibinfo{author}{\bibfnamefont{J.~J.} \bibnamefont{Ling}},
  \bibinfo{author}{\bibfnamefont{Y.}~\bibnamefont{Nakajima}}, \bibnamefont{and}
  \bibinfo{author}{\bibfnamefont{C.}~\bibnamefont{Zhang}}, ``{The Gaussian
  CL$_s$ method for searches of new physics},'' \bibinfo{journal}{Nucl.
  Instrum. Meth. A} \textbf{\bibinfo{volume}{827}}, \bibinfo{pages}{63}
  (\bibinfo{year}{2016}), \eprint{1407.5052}.

\bibitem[{\citenamefont{Feldman and Cousins}(1998)}]{Feldman:1997qc}
\bibinfo{author}{\bibfnamefont{G.~J.} \bibnamefont{Feldman}} \bibnamefont{and}
  \bibinfo{author}{\bibfnamefont{R.~D.} \bibnamefont{Cousins}}, ``{A Unified
  approach to the classical statistical analysis of small signals},''
  \bibinfo{journal}{Phys. Rev.} \textbf{\bibinfo{volume}{D57}},
  \bibinfo{pages}{3873} (\bibinfo{year}{1998}), \eprint{physics/9711021}.

\bibitem[{\citenamefont{Agnolet et~al.}(2017)}]{Agnolet:2016zir}
\bibinfo{author}{\bibfnamefont{G.}~\bibnamefont{Agnolet}} \bibnamefont{et~al.}
  (\bibinfo{collaboration}{MINER}), ``{Background Studies for the MINER
  Coherent Neutrino Scattering Reactor Experiment},'' \bibinfo{journal}{Nucl.
  Instrum. Meth. A} \textbf{\bibinfo{volume}{853}}, \bibinfo{pages}{53}
  (\bibinfo{year}{2017}), \eprint{1609.02066}.

\bibitem[{\citenamefont{Danilov}(2020)}]{Danilov:2019aef}
\bibinfo{author}{\bibfnamefont{M.}~\bibnamefont{Danilov}}
  (\bibinfo{collaboration}{DANSS}), ``{Recent results of the DANSS
  experiment},'' \bibinfo{journal}{PoS} \textbf{\bibinfo{volume}{EPS-HEP2019}},
  \bibinfo{pages}{401} (\bibinfo{year}{2020}), \eprint{1911.10140}.

\bibitem[{\citenamefont{Serebrov}()}]{Neutrino4talk}
\bibinfo{author}{\bibfnamefont{A.~P.} \bibnamefont{Serebrov}}, ``{Present
  status of Neutrino-4 experiment search for sterile neutrino},'',
  \urlprefix\url{https://indico.cern.ch/event/833568/contributions/3655173/attachments/1957823/3252790/2-12_China_Serebrov_Neutrino-4.pdf}.

\bibitem[{\citenamefont{Aker et~al.}(2021)}]{KATRIN:2020dpx}
\bibinfo{author}{\bibfnamefont{M.}~\bibnamefont{Aker}} \bibnamefont{et~al.}
  (\bibinfo{collaboration}{KATRIN}), ``{Bound on 3+1 Active-Sterile Neutrino
  Mixing from the First Four-Week Science Run of KATRIN},''
  \bibinfo{journal}{Phys. Rev. Lett.} \textbf{\bibinfo{volume}{126}},
  \bibinfo{pages}{091803} (\bibinfo{year}{2021}), \eprint{2011.05087}.

\bibitem[{\citenamefont{Formaggio et~al.}(2021)\citenamefont{Formaggio,
  de~Gouv\^ea, and Robertson}}]{Formaggio:2021nfz}
\bibinfo{author}{\bibfnamefont{J.~A.} \bibnamefont{Formaggio}},
  \bibinfo{author}{\bibfnamefont{A.~L.~C.} \bibnamefont{de~Gouv\^ea}},
  \bibnamefont{and} \bibinfo{author}{\bibfnamefont{R.~G.~H.}
  \bibnamefont{Robertson}}, ``{Direct Measurements of Neutrino Mass},''
  \bibinfo{journal}{Phys. Rept.} \textbf{\bibinfo{volume}{914}},
  \bibinfo{pages}{1} (\bibinfo{year}{2021}), \eprint{2102.00594}.

\bibitem[{\citenamefont{Bergevin et~al.}(2019)}]{Bergevin:2019tcg}
\bibinfo{author}{\bibfnamefont{M.}~\bibnamefont{Bergevin}}
  \bibnamefont{et~al.}, ``{Applied Antineutrino Physics 2018 Proceedings},''
  (\bibinfo{year}{2019}), \eprint{1911.06834}.

\bibitem[{\citenamefont{Lyashuk}(2016)}]{Lyashuk:2016lpn}
\bibinfo{author}{\bibfnamefont{V.~I.} \bibnamefont{Lyashuk}}, ``{High flux
  lithium antineutrino source with variable hard spectrum. How to decrease the
  errors of the total spectrum?},''  (\bibinfo{year}{2016}),
  \eprint{1612.08096}.

\bibitem[{\citenamefont{Alonso and Nakamura}(2017)}]{Alonso:2017fci}
\bibinfo{author}{\bibfnamefont{J.~R.} \bibnamefont{Alonso}} \bibnamefont{and}
  \bibinfo{author}{\bibfnamefont{K.}~\bibnamefont{Nakamura}}
  (\bibinfo{collaboration}{IsoDAR}), ``{IsoDAR@KamLAND: A Conceptual Design
  Report for the Conventional Facilities},''  (\bibinfo{year}{2017}),
  \eprint{1710.09325}.

\bibitem[{\citenamefont{Conant}(2018)}]{conant}
\bibinfo{author}{\bibfnamefont{A.}~\bibnamefont{Conant}}, Ph.D. thesis,
  \bibinfo{school}{Georgia Tech} (\bibinfo{year}{2018}).

\end{thebibliography}

\end{document}